\documentclass[journal]{IEEEtran}
\usepackage{amsfonts}
\usepackage{amsmath}
\usepackage{amssymb}
\usepackage{tabularx}
\usepackage{tikz}
\usepackage{array}
\usepackage{textcomp}
\usepackage{stfloats}
\usepackage{url}
\usepackage{verbatim}
\usepackage{graphicx}
\usepackage{subcaption}
\usepackage{orcidlink}
\usepackage{cleveref}
\usepackage{diagbox}
\usepackage{colortbl}
\usepackage{xcolor}
\def\BibTeX{{\rm B\kern-.05em{\sc i\kern-.025em b}\kern-.08em
		T\kern-.1667em\lower.7ex\hbox{E}\kern-.125emX}}
\usepackage{balance}
\usepackage{algorithm}
\usepackage{algorithmic}
\usepackage{cite}
\usepackage{makecell}
\usepackage{esint}
\usepackage{tabularray}
\usepackage{dsfont}
\usepackage{booktabs}

\usepackage{soul}
\usepackage{esvect}
\usepackage{mathtools}
\usepackage{physics}
\usepackage{gensymb}
\usepackage{tablefootnote}
\usepackage{cleveref}

\usepackage{bm}

\usepackage{soul}

\usetikzlibrary{calc,arrows.meta,positioning}

\newcommand{\ie}{, i.e., }
\newcommand{\eg}{, e.g., }

\crefformat{equation}{~(#2#1#3)}
\crefname{figure}{Fig.}{Figs.} 

\crefrangeformat{equation}{~(#3#1#4)-(#5#2#6)}

\begin{document}
	\title{MPSi: Neural Extraction of Site-Specific Multipath Propagation Properties for Massive MIMO}
	\author{
	Maryam Rezvani\orcidlink{https://orcid.org/0000-0002-4492-2547}, \textit{Graduate Student Member, IEEE}, 
	Raviraj Adve\orcidlink{https://orcid.org/0000-0003-0224-2209}, \textit{Fellow, IEEE}, 
	Amr El-Keyi\orcidlink{https://orcid.org/0000-0003-2903-4055}, \textit{Member, IEEE}, 
	and Akram bin Sediq\orcidlink{https://orcid.org/0000-0003-1260-2853}
		\thanks{
			Manuscript received November 2025; revised XXX XXX; accepted XXX XXXX. Date of publication XXX; date of current version XXXX XXX. The associate editor coordinating the review of the manuscript and approving it for publication was XXX. This work was supported by Ericsson Canada.\\ \indent 
			Maryam Rezvani and Raviraj Adve are with the Department of Electrical and Computer Engineering, University of Toronto, Toronto, Ontario M5S 3G4, Canada (e-mail: mrezvani@ece.utoronto.ca; rsadve@comm.utoronto.ca). \\  
			Akram bin Sediq and Amr El-Keyi are with Ericsson Canada, Kanata, ON K2K 2V6, Canada (e-mail: akram.bin.sediq@ericsson.com, amr.el-keyi@ericsson.com). \\ \indent 
			Contact mrezvani@ece.utoronto.ca for further questions about this work.\\ \indent 
			Digital Object Identifier 
	}
	}
	\markboth{IEEE Transactions on Wireless Communications,~Vol.~XX, No.~X, XXX~2026}%
	{Fast Initial Access in Massive MIMO via Deep‑Unfolded Channel Sounding}

	\maketitle
	\begin{abstract}
	Accurate extraction of site-specific multipath propagation properties enables a wide range of wireless applications, including initial access, wideband channel reconstruction, and the generation of channel datasets for learning-based and continual-learning systems. Classical parametric channel-sounding algorithms such as RIMAX provide physically interpretable estimates, but their data-dependent model-order selection and iterative nonlinear refinement introduce substantial delay and computational cost. This paper proposes MPSi, a physics-guided neural framework that replaces these costly stages with a compact path-count network and a fixed-depth, deep-unfolded Levenberg–Marquardt refinement. MPSi estimates the complex gain, angle of arrival, and propagation delay of each multipath component while preserving the underlying physical channel model. We evaluate MPSi in two site-specific ray-traced environments, a small cell and a dense urban area, and show that it achieves initial-access and full-channel-estimation performance comparable to an enhanced RIMAX baseline while requiring significantly fewer optimization iterations. The reconstructed channels also yield bit-error-rate performance close to the perfect-channel knowledge reference. These results position MPSi as a promising low-overhead framework for extracting multipath propagation properties in massive-MIMO systems and for generating cell-specific channel pseudo-labels for training when direct ground-truth measurements are unavailable.
	\end{abstract}
	
	
	\section{Introduction}
	\IEEEPARstart{M}{odern} wireless systems increasingly rely on accurate knowledge of the underlying communication channel. Physically, the channel is composed of multiple propagation paths whose delays, angles of arrival (AoAs), and complex gains combine to form the observed signal. Different applications in massive multiple-input multiple-output (MIMO) systems require different subsets of this electromagnetic structure: initial access primarily depends on the dominant-path AoA and is typically handled through codebook-based beam searches~\cite{HeathCodebook,Codebook}, whereas wideband channel acquisition requires the full set of path parameters to reconstruct the channel across antennas and frequencies~\cite{minn2000investigation}. 
	
	With the rise of machine-learning (ML)–based algorithms in wireless communication, the need for accurate channel representations and reliable data generation has become even more critical to train these models effectively~\cite{MLsurvey}. If a single technique can accurately recover the true multipath structure of the environment, then it can simultaneously support all of these applications. This motivates the search for a unified, physically grounded estimator of the channel’s electromagnetic structure.
	
	A variety of techniques currently address these applications individually. For initial access, codebook-based beam searches are commonly used~\cite{HeathCodebook,Codebook}, where multiple beams are probed to identify the	strongest initial direction. This procedure introduces delay and prevents the user from quickly operating at full link capacity. For channel acquisition, the complete set of path parameters enables reconstruction over unmeasured subcarriers~\cite{minn2000investigation}, whereas conventional estimators must compute the channel on every subcarrier, increasing both pilot overhead and computational complexity. 
	
	A parametric time-domain model captures the underlying electromagnetic structure using the delay, AoA, and complex gain of each propagation path~\cite{TDDCE,minn2000investigation,mirzaei2018semi}, reducing the dimensionality of the estimation problem while preserving the physical relationships governing how the channel varies across antennas and frequency. A method that extracts this multipath structure from limited pilot observations would therefore serve as a unified physical-layer tool for both initial access and channel estimation.
	
	Separately, a major obstacle in the design of learning-base methods, is obtaining the training data needed for site-specific ML methods. Many ML-enhanced signal-processing frameworks, including learned codebook designs and learned channel estimation, depend on channel state information or other idealized labels as training targets~\cite{HeathCodebook, mirzaei2022deep,physicsNn1, MLsurvey}. In practice, extracting these targets purely from field measurements remains challenging. When physical channel data is unavailable, existing studies typically rely on offline ray-tracing simulations or synthetic generative tools, leaving deployment-phase label acquisition unresolved. This bottleneck affects all site-specific ML applications and becomes even more severe for continual-learning systems~\cite{CLCioff,CLearning} that require a real-time stream of fresh ground-truth labels.
	
	While site-specific ray tracing can provide the electromagnetic properties of the communication channel needed for these applications, it incurs prohibitive computational complexity and execution  delays~\cite{raytracing}, preventing its use as an online engine for automated retraining pipelines or real-time wireless operation. Neural-network accelerations have been proposed, but they either rely
	on an integrated ray-tracing engine to supply fine-grained path parameters such as AoAs on the fly~\cite{winert}, or restrict their scope to predicting macro-level features like path loss from environmental maps
	\cite{seretis2020ray,seretis2022toward}. Physics-inspired neural networks embed the propagation model through latent parameters, but they do not explicitly isolate individual active paths; instead, they learn the aggregate geometric effect~\cite{physicsNn1}, producing accurate channel estimates but yielding uninterpretable path parameters that cannot be directly mapped to other applications, including initial access and label generation.
	
	Classical physics-based channel-sounding methods such as SAGE~\cite{sage} and RIMAX~\cite{rimax} can recover multipath parameters accurately using the known sounding sequence, transmit power, and received signal. RiMAX introduced as a more accurate replacement for SAGE, combines data-dependent path-count selection, diffuse-multipath component (DMC) modeling, and iterative nonlinear refinement with cost-based step control. These operations yield high-fidelity estimates but introduce variable control flow and an unknown number of repeated optimization steps, making RIMAX unsuitable for low-delay wireless systems. Therefore, a predictable-depth technique that can extract realistic active-path parameters would be valuable for practical wireless communication.
	
	Motivated by these limitations, we take a gray-box approach~\cite{bohlin2006practical,youssry2026bayesian} that integrates a neural network with a physics-based optimization framework. The proposed method uses deep unfolding and a small fully connected network to reduce the data-dependent complexity of RIMAX while preserving its physical under pinning. The network predicts an effective number of propagation paths, find the initial value for each path parameter, and uses a fixed number of unfolded Levenberg--Marquardt (LM) layers to refine the initial values. This yields a predictable-depth estimator that extracts the delay, AoA, and complex gain of the multipath components from a single pilot observation. The resulting estimates serve both as fixed-depth channel-sounding outputs and as auditable candidate labels for wireless-learning systems when perfect channel knowledge is unavailable.\footnote{Doppler shift can be added as another path parameter in future extensions for channel prediction~\cite{channelPrediction} and further pilot-overhead reduction~\cite{rimax}.}
	
	We refer to the proposed method as \emph{Multi-Path Sensing integrated with neural networks} (MPSi).
	
	\textbf{Contributions:}	The main contributions of this work are summarized as follows:
	\begin{itemize}
		\item We introduce MPSi, a deep-unfolded Levenberg--Marquardt (LM) architecture
		that estimates geometric path parameters from the received pilot signal,
		the known sounding sequence, and the pilot power.
		
		\item We develop an unfolded LM update with trainable damping, a trainable step-trust schedule, and a Fisher-information-matrix-related positive-definite floor. These controls were introduced to prevent the early training instability observed with unconstrained unfolded updates, while replacing the nondifferentiable accept/reject loop of classical LM with stable, fully differentiable updates.
		
		\item We propose a lightweight path-count network and a differentiable soft-top-$k$ gate that assigns relevance weights to a fixed set of initialized path candidates. This differentiable path-count mechanism allows model-order information to be learned jointly with the physical channel parameters.
		
		\item We train the complete architecture end to end using a reconstruction objective that combines pilot-domain normalized mean-squared error, deep supervision across LM layers, and a path-count parsimony penalty.
		
		\item We construct training, validation, and test datasets using the Sionna ray tracer~\cite{sionna} and 3D environments derived from OpenStreetMap data~\cite{openstreetmap}, with basestation (BS) locations and heights selected from operator deployment maps.
		
		\item We evaluate MPSi for initial access, channel estimation, and pseudo‑label generation, showing its versatility across multiple wireless‑communication tasks.
		
		\item We release the implementation of MPSi, including the trained models and dataset-generation scripts, as open-source resources to facilitate reproducibility and further research\footnote{\url{https://github.com/R-Maryam/MPSi}}.
	\end{itemize}
	
	\textit{Organization:} The remainder of this paper is organized as follows. Section~\ref{sec:systemModel} presents the system model, including the multipath channel representation and the pilot observation model. Section~\ref{sec:PS} formulates the channel parameter extraction problem and reviews the classical RIMAX baseline. Section~\ref{sec:MPSi} introduces the proposed MPSi architecture, its training objective, and the dataset-generation procedure. Section~\ref{sec:simulations} presents the simulation results and compares MPSi with the considered baselines. Finally, Section~\ref{sec:conclusion} concludes the paper and discusses directions for future work.
	
	\textit{Notation:} We use lowercase and uppercase boldface letters to denote vectors and matrices, respectively. The operators $(\cdot)^H$, $(\cdot)^{-1}$, and $\|\cdot\|_2$ denote the conjugate transpose, matrix inverse, and second-order norm, respectively. The operators $\Re\{\cdot\}$ and $\Im\{\cdot\}$ denote the real and imaginary parts, and $\mathbb{E}[\cdot]$ denotes statistical expectation. We use $\jmath=\sqrt{-1}$, $\mathbb{C}$ for the set of complex numbers, and $\mathbb{R}$ for the set of real numbers. The notation $\mathbf{A}_{n,m}$ denotes the element in the $n$th row and $m$th column of matrix $\mathbf{A}$, while $[\mathbf{a}]_n$ denotes the $n$th element of vector $\mathbf{a}$. We use $\mathcal{CN}(\mu,\sigma)$ to denote a complex Gaussian distribution with mean $\mu$ and covariance $\sigma$.
	
	\section{System Model}\label{sec:systemModel}
	We consider an uplink system in which the BS employs a uniform planar array (UPA) with $N_x$ horizontal and $N_y$ vertical antenna elements, for a total of $N=N_xN_y$ antennas. The user transmits pilot symbols over $M$ equispaced measured subcarriers out of a total of $N_{scr}$ subcarriers of an orthogonal frequency division multiplexing/multiple access (OFDM/OFDMA) symbol. Users are assumed to be stationary and equipped with a single antenna.
	
	
	The channel is modeled as a superposition of the propagation paths between the user and the BS array~\cite{rimax}:
	\begin{equation}
		\mathbf{H}_{n,m} = \sum_{p=1}^P \alpha_p [\mathbf{a}(\theta_p, \phi_p)]_ne^{-\jmath 2\pi \nu_m \tau_p}, \label{eq:channelmodel}
	\end{equation}	
	where $\mathbf{H}_{n,m}$ is the channel coefficient at antenna $n$ and measured subcarrier $m$, $P$ is the number of resolvable paths, $\alpha_p$ is the complex baseband gain (which absorbs the carrier-frequency phase), $\theta_p$ and $\phi_p$ are the elevation and azimuth AoAs, $\tau_p$ is the path delay, and $\nu_m$ is the signed baseband frequency offset of FFT bin $m$. The UPA steering vector is $\mathbf{a}=\mathbf{a}_y\otimes\mathbf{a}_x$, with
	\begin{equation}
		\mathbf{a}_{c}(\theta_p,\phi_p) = \left [1, e^{\jmath  u_c(\theta_p,\phi_p)}, \ldots, e^{\jmath (N_c-1) u_c(\theta_p,\phi_p)}\right]^T,
	\end{equation}
	where $\mathbf{u}(\theta_p,\phi_p)=2 \pi \tfrac{d_c}{\lambda} \dot[\cos\theta_p \cos\phi_p,\cos\theta_p \sin\phi_p,\sin\theta_p]^T$ is the propagation-direction vector, $u_c$ is its component along array axis $c\in\{x,y\}$, $N_c$ is the number of elements along that axis, $d_c$ is the corresponding inter-element spacing, and $k=2\pi/\lambda$ is the wavenumber. Here, $\theta_p$ is measured from the horizontal plane and $\phi_p\in[-\pi,\pi]$ is the azimuth.
	
	We use a constant-modulus quadratic-phase sounding sequence as transmitted pilots~\cite{rimax}. 
	For a pilot length $M$ and root index $w$, the sounding sequence is defined as:
	\begin{equation}
		\mathbf{s}_{\mathrm{pilot}}[m] =	\exp\left(-\jmath\frac{\pi w m(m+1)}{M}\right),\ m=0,\ldots,M-1.
	\end{equation}
	The sequence has unit average power and its discrete Fourier transform (DFT) results in the frequency-domain pilot $\mathbf{s}_{\mathrm{pilot}}[m]$.
	
	The resulting baseband frequency-domain pilot observation at the BS, $\mathbf{Y}_\mathrm{pilot}\in\mathbb{C}^{N\times M}$, is
	\begin{equation}
		\mathbf{Y}_{\mathrm{pilot}}[n,m] = \sqrt{P_T} \mathbf{H}_{n,m}\mathbf{s}_{\mathrm{pilot}}[m] + \mathbf{W}[n,m].
		\label{eq:systemModel}
	\end{equation}
	Here, $P_T$ is the pilot transmit power and $\mathbf{W}[n,m]$ is the circularly symmetric complex Gaussian receiver-noise sample with variance $\sigma_n^2$. We emphasize that the $M$ pilot measurements are collected on frequency-domain subcarriers within a single time slot and a single OFDM/OFDMA symbol.
	
	Our goal is to estimate the path set \(\{\boldsymbol{\Phi}_p\}_{p=1}^{P}=\{(\alpha_p,\theta_p,\phi_p,\tau_p)\}_{p=1}^{P}\). These parameters capture the electromagnetic properties of the communication
	channel and enable several key functions: identifying the dominant arrival direction for beam alignment, providing a low-dimensional representation from which the full spatial–frequency channel \(\mathbf{H}\) can be reconstructed, and generating pseudo-labels for downstream ML algorithms.
	
	\section{Problem Statement}\label{sec:PS}
	Given $\mathbf{Y}_\mathrm{pilot}$, our goal is to estimate the number and parameters of the dominant propagation paths\ie solving the following optimization problem:
	\begin{subequations}
		\label{eq:ProbStatement}
		\begin{align}
			&\min_{P, \{\mathbf{\Phi}_p\}_{p=1}^{P}} \sum_{m=0}^{M-1} \nonumber\\ &\norm{\mathbf{Y}_{\mathrm{pilot}}[:,m] -\sqrt{P_T}S_{\mathrm{pilot}}[m]\sum_{p=1}^{P}\alpha_p\mathbf{a}(\phi_p,\theta_p)e^{-\jmath 2\pi \nu_m \tau_p}}_2^2 \label{eq:PathEstimation_obj}\\
			&P \in \{P_\mathrm{min}, \cdots, P_\mathrm{max}\}, \label{eq:PathEstimation_pathnum} \\
			&\tau_p \in [0,\tau_{\max}]. \label{eq:PathEstimation_tau}
		\end{align}
	\end{subequations}
	The objective in~\eqref{eq:ProbStatement} jointly estimates the active path count and the corresponding gains, delays, and AoAs by minimizing the pilot-reconstruction error. Constraint~\eqref{eq:PathEstimation_pathnum} bounds the model order and~\eqref{eq:PathEstimation_tau}-\eqref{eq:PathEstimation_tau} specify the physical parameter domains.
	
	Measuring in the frequency domain gives us the opportunity to minimize the delay associated with collecting the data points needed for path-parameter extraction. On the other hand, classical channel-sounding algorithms such as SAGE~\cite{sage} and RIMAX~\cite{rimax} operate in the time domain. To leverage their insights and algorithmic benefits, we therefore transform the optimization problem in~\eqref{eq:ProbStatement} into the time domain as well. The advantages of this transformation will become clearer in the next section. Accordingly, we apply an inverse discrete Fourier transform (IDFT) to the \(M\) frequency-domain measurements and form the equivalent time-domain objective as
	\begin{subequations}
		\label{eq:ProbStatementTD}
		\begin{align}
			&\min_{P,\{\mathbf{\Phi}_p\}_{p=1}^{P}}
			\sum_{m=0}^{M-1}\nonumber\\
			&\left\|
			\mathbf{Y}_{\mathrm{td}}[:,m]
			-
			\sqrt{P_T}\sum_{p=1}^{P}
			\alpha_p\,
			\mathbf{a}(\phi_p,\theta_p)\,
			\mathbf{s}_{\mathrm{td}}(m - \tau_p f_s)
			\right\|_2^2,
			\label{eq:PathEstimationTD_obj}\\
			&\eqref{eq:PathEstimation_pathnum} - \eqref{eq:PathEstimation_tau},
		\end{align}
	\end{subequations}
	where $\mathbf{Y}_{\mathrm{td}}\in\mathbb{C}^{N\times M}$ is the time domain representation of the measured pilot matrix, $f_s$ is the IDFT sampling rate equal to system bandwidth, and $\tau_p f_s$ represents the path delay expressed in samples.
	
	The main challenge in applying SAGE and RIMAX to solve~\eqref{eq:ProbStatementTD} is their computational complexity. SAGE~\cite{sage} requires an \textit{a priori} path count and performs iterative expectation–maximization updates. Each iteration visits the paths sequentially and refines their parameters while holding the remaining estimates fixed, resulting in data‑dependent processing time and strict sequential dependencies.
	
	RIMAX~\cite{rimax}, by contrast, was designed to achieve higher‑accuracy estimation than SAGE by performing joint maximum‑likelihood refinement using a prune‑and‑add model‑order strategy combined with an LM nonlinear optimization step. RIMAX operates through an outer loop that manages the model order, where paths are added or pruned based on stability and contribution criteria. Within each outer iteration, an inner LM loop jointly refines all path parameters. The LM update employs a trust‑region mechanism to select the step size, evaluating multiple candidate step lengths to prevent divergence and reduce the likelihood of converging to poor local optima. This combination of joint parameter updates, trust‑region step‑size selection, and dynamic model‑order adaptation leads to high computational complexity and long execution times.
	
	It is worth noting that because the objective in~\eqref{eq:PathEstimationTD_obj} is nonconvex, all algorithms are sensitive to initialization. We will discuss this issue in Section~\ref{sec:initializer}.
	
	This high complexity makes these classical algorithms unsuitable for the real‑time demands of modern wireless‑communication applications. We therefore integrate neural‑network and machine‑learning techniques with the algorithmic insights of RIMAX to obtain a high‑accuracy, low‑complexity estimator for solving~\eqref{eq:ProbStatementTD}.
	
	\section{MPSi: Physics-Guided Neural Multipath Sensing}\label{sec:MPSi}
\begin{figure}[t]
		\centering
		\resizebox{0.85\columnwidth}{!}{%
		\begin{tikzpicture}[
			>=Latex,
			node distance=6mm and 3mm,
			block/.style={
				draw,
				rounded corners=2pt,
				align=center,
				minimum height=9mm,
				text width=30mm,
				font=\footnotesize
			},
			wideblock/.style={
				draw,
				rounded corners=2pt,
				align=center,
				minimum height=9mm,
				text width=28mm,
				font=\footnotesize
			},
			arrow/.style={->,thick},
			trainarrow/.style={->,thick,dashed}
		]
			\node[block] (pilot) {Received pilot\\$\mathbf{Y}_{\mathrm{pilot}}\in\mathbb{C}^{N\times M}$};
			\node[block, below=of pilot] (preprocess) {Power normalization\\and IFFT\\$\mathbf{Y}_{\mathrm{td}}$};
			\node[wideblock, below left=7mm and 0mm of preprocess] (pcn) {Path Count Network\\$P_{\mathrm{est}}\in[P_{\min},P_{\max}]$};
			\node[wideblock, below right=7mm and 0mm of preprocess] (init) {Initialization Method\\$P_{\max}$ candidate paths and initial-gain ranking};
			\node[block, below=28mm of preprocess] (gate) {Soft-top-$K$ gate\\$\mathbf{g}$};
			\node[wideblock, below=of gate] (lm) {Fixed-depth unfolded LM\\ridge-LS gains, learned damping,\\and learned step trust};
			\node[block, below=of lm] (state) {Refined candidate state\\$\{\alpha_p,\phi_p,\theta_p,\tau_p\}$};
			\node[wideblock, below left=7mm and 0mm of state] (training) {Training\\gated reconstruction, deep supervision, and count penalty};
			\node[wideblock, below right=7mm and 0mm of state] (inference) {Inference and applications\\hard gating and gain refit; initial access, channel reconstruction and detection, and pseudo-labels};

			\draw[arrow] (pilot) -- (preprocess);
			\draw[arrow] (preprocess) -- (pcn);
			\draw[arrow] (preprocess) -- (init);
			\draw[arrow] (pcn) -- (gate);
			\draw[arrow] (init) -- (gate);
			\draw[arrow] (gate) -- (lm);
			\draw[arrow] (lm) -- (state);
			\draw[trainarrow] (state) -- (training);
			\draw[arrow] (state) -- (inference);
		\end{tikzpicture}
		}
		\caption{Overall MPSi processing flow.}
		\label{fig:mpsi-overview}
	\end{figure}
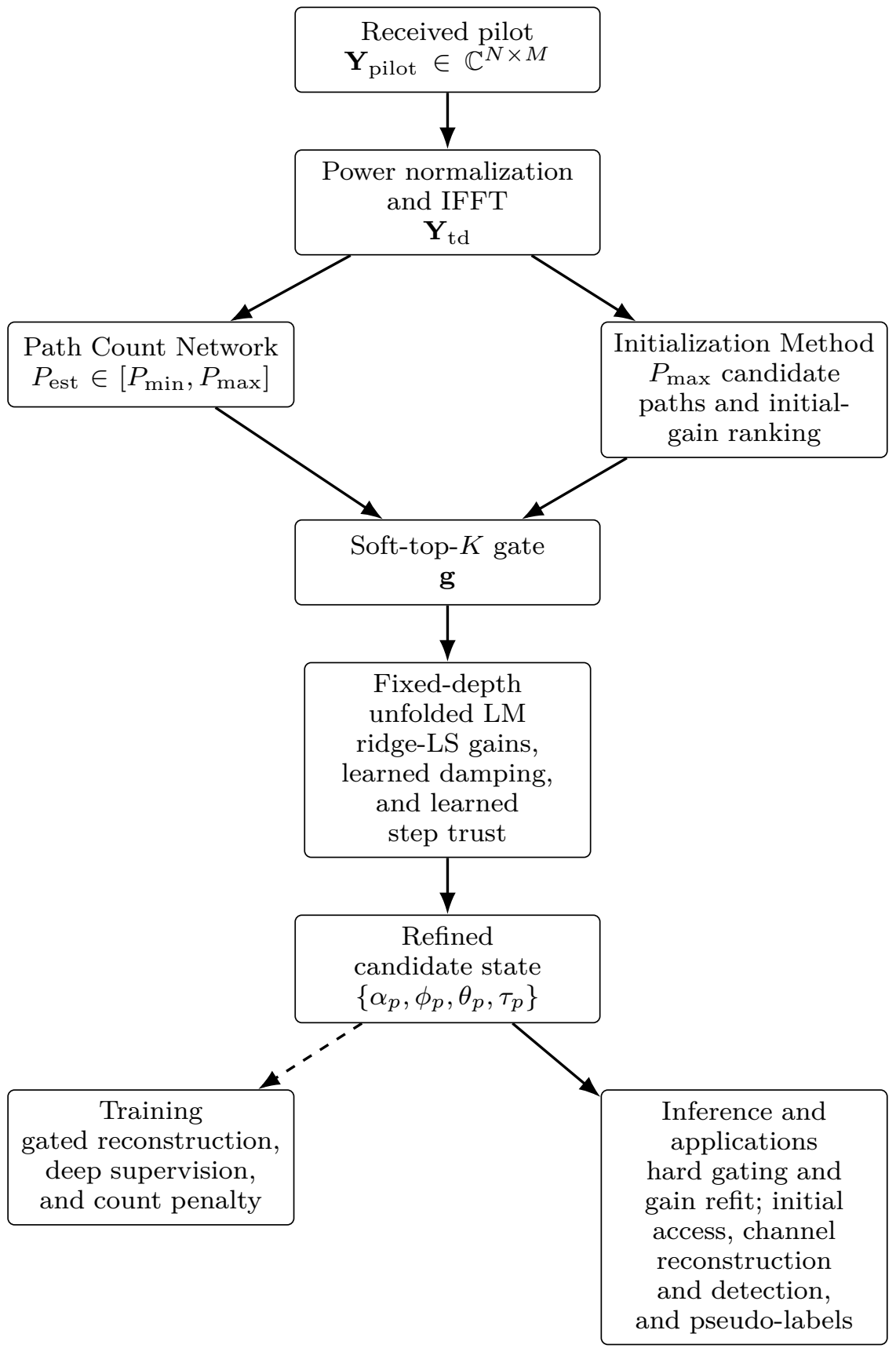

	Our proposed technique, MPSi, integrates compact neural components with a physics-based parametric channel model to extract site-specific multipath properties from received pilot measurements. As summarized in Fig.~\ref{fig:mpsi-overview}, at a high level, the framework first converts the transmit-power-normalized frequency-domain pilot into a time-domain sounding observation. A neural network, the PCN, then predicts an effective path count. In parallel, the initializer constructs a fixed set of \(P_{\max}\) candidate
	paths whose initial gains determine their ranking. The predicted count and candidate ranking together form differentiable soft-top-\(K\) gates. Finally, a fixed-depth unfolded LM refiner updates the candidate geometries and computes ridge-regularized~\cite{ridge} joint gain estimates, while the gates attenuate candidate contributions during reconstruction. The following sections detail each component of this design.

	\subsection{Path Count Network}\label{sec:PCN}
	
	The first learned component of MPSi is the path-count network (PCN). Given a fixed maximum number of candidate paths \(P_{\max}\), the PCN predicts the effective number of active paths. During training, this continuous prediction controls a differentiable gate, while during inference, thresholding the gate yields a discrete retained set. This mechanism provides a soft relaxation of top-\(K\) selection and is related to differentiable ranking and sorting methods~\cite{TopK,NeuralSort,SoftSort}.
	
	The PCN input is formed using an antenna-averaged power profile computed as
	\begin{equation}
		P_{\mathrm{avg}}^{m} = \frac{1}{N} \sum_{n=1}^{N} P_{\mathrm{obs}}^{(n,m)},
	\end{equation}
	where \(P_{\mathrm{obs}}^{(n,m)} = |\widetilde{Y}_{\mathrm{pilot}}^{(n,m)}|^{2}\) denotes the normalized observed power at measurement index \(m\) and antenna \(n\), and
	\(\widetilde{\mathbf{Y}}_{\mathrm{pilot}} = \mathbf{Y}_{\mathrm{pilot}}/\sqrt{P_T}\).
	This representation is closely related to the power-delay profile, whose shape, spread, and dominant peaks reflect the richness of the underlying multipath environment~\cite{ITU-PDP}. Averaging across antennas substantially reduces the input dimension while preserving the large-scale multipath structure. The PCN is therefore trained to infer the effective number of resolvable paths from this power distribution.
	
	To enhance the distinguishability between paths with similar power levels, we compute a logarithmic power feature, which serves as the input to the PCN:
	\begin{equation}
		\mathbf{q}_{m} = \log\!\left( P_{\mathrm{avg}}^{m} + \epsilon \right).
	\end{equation}
	The small constant $\epsilon$ ensures numerical stability when $P_{\mathrm{avg}}^{m}$ approaches zero. The resulting feature vector $\mathbf{q}$ is then passed through the PCN, which consists of $\ell_{\mathrm{PCN}}$ fully connected layers with ReLU activations. The $i$th hidden layer computes
	\begin{equation}
		\mathbf{f}_i = \mathrm{ReLU}\!\left( \mathbf{W}_i \mathbf{f}_{i-1} + \mathbf{b}_i \right),
	\end{equation}
	where $\mathbf{W}_i$ and $\mathbf{b}_i$ are the weight matrix and bias of the $i$th layer, and $\mathbf{f}_{i-1}$ is the output of the $(i-1)$th layer (with $\mathbf{f}_0 = \mathbf{q}$).
	
	The output layer of the PCN uses a sigmoid activation, $\sigma(x)=\tfrac{1}{1+e^{-x}}$, to produce a continuous estimate of the number of paths between the predetermined bounds:
	\begin{equation}
		P_{\mathrm{est}} = P_{\mathrm{min}} + (P_{\mathrm{max}} - P_{\mathrm{min}})\,
		\sigma\!\left( \mathbf{W}_{\mathrm{out}} \mathbf{f}_{\ell_{\mathrm{PCN}}} + \mathbf{b}_{\mathrm{out}} \right).
	\end{equation}
	Thus, $P_{\mathrm{est}}$ is a continuous scalar in $[P_{\mathrm{min}},P_{\mathrm{max}}]$ and is not rounded during training.
	
	\textbf{Differentiable soft-top-$K$ path gate:} Rather than pruning candidates during training, MPSi assigns each initialized path a continuous activation weight $\mathbf{g}_p\in(0,1)$.
	
	The initial values of all path gains (obtained as discussed in Section~\ref{sec:initializer}) are used to determine the ordering of the candidate paths. After the initializer computes the gains $\alpha_p$ (Section~\ref{sec:initializer}), their magnitudes define the rank $\rho_p$ as
	\begin{equation}
		\rho_p = 1 + \mathtt{argsort}\!\left(-|\mathbf{\alpha}|\right)_p,
	\end{equation}
	where $\mathtt{argsort}$ returns the indices that sort its argument in ascending order. Note that the ranking is used only to establish the ordering of the candidate paths and is treated as a constant during training. Consequently, gradients do not propagate through the $\mathtt{argsort}$ operation.
	
	Given the continuous path-count prediction $P_{\mathrm{est}}$, the differentiable soft-top-$K$ gate is defined as
	\begin{equation}
		\mathbf{g}_p = \sigma\!\left( \frac{P_{\mathrm{est}} - \mathbf{\rho}_p}{T} \right),
		\label{eq:gate}
	\end{equation}
	where $T$ is a temperature parameter controlling the sharpness of the gating transition. When $\mathbf{\rho}_p < P_{\mathrm{est}}$, the argument of the sigmoid in~\eqref{eq:gate} is positive, yielding $\mathbf{g}_p \approx 1$. Conversely, when $\mathbf{\rho}_p > P_{\mathrm{est}}$, the argument becomes negative and $\mathbf{g}_p \approx 0$.
	
	During training, the gate values weight the candidate contributions in the channel model:
	\begin{equation}
		\mathbf{H}_{n,m}^{g}
		= \sum_{p=1}^{P_{\max}} \mathbf{g}_p \alpha_p \mathbf{a}_n(\theta_p, \phi_p)e^{-\jmath 2\pi \nu_m \tau_p},
		\label{eq:channelmodelGate}
	\end{equation}
	where $\mathbf{H}_{n,m}^{g}$ is the gated channel response at antenna $n$ and subcarrier $m$. Incorporating the gate into the channel model allows candidate paths to be softly selected during training while preserving end-to-end differentiability. 
	
	Note that since the sigmoid gate is differentiable with respect to $P_{\mathrm{est}}$, the reconstruction loss and its gradient propagate through the gate to the PCN. Consequently, the network learns the number of paths that contribute to the reconstruction of the communication channel. At inference time, this differentiable mechanism is replaced by a hard selection of the active paths, defined as $\mathcal{P} = \{ p : \mathbf{g}_p \geq \gamma_{\mathrm{gate}}\}$, where $\gamma_{\mathrm{gate}}=0.5$.

	
	Note that the RIMAX and MPSi path-selection procedures differ fundamentally. RIMAX distinguishes specular multipath components from a dense-multipath component that is represented statistically through a covariance model~\cite{rimax}. Its model-order selection relies on iterative prune-and-add operations that evaluate whether newly introduced components reduce the cost function, adding to the algorithm’s complexity and data-dependent control flow. By contrast, MPSi operates on a finite candidate set and uses a learned gate to retain the components that improve reconstruction under the adopted discrete-path model. Accordingly, the estimated count should be interpreted as the number of effective \textit{resolvable} components supported by the model and measurement configuration, rather than as a direct count of all physical scattering mechanisms.
	
	In the next section, we detail the procedure used to compute the initial path parameters.
	
	\subsection{Initialization}\label{sec:initializer}
	
	The objective in~\eqref{eq:PathEstimationTD_obj} is nonconvex in both delay and AoA, making initialization critical for the accuracy of the subsequent refinement. We adopt RIMAX's successive multipath initialization with a coarse angular grid and augment it with an additional Nelder–Mead (NM) refinement step\footnote{NM is a derivative-free optimization method similar to pattern search; it dynamically adapts a line or a two-dimensional triangle to locate the function's minimizer.}~\cite{nelder1965simplex}. Delay is refined through a one-dimensional search, while the angular update performs a two-dimensional NM search over azimuth and elevation. The resulting initialized angles are then converted to the scaled spatial–frequency coordinates used by the remainder of the MPSi algorithm.
	
	The frequency-domain observation is first normalized by the known pilot-amplitude scale and transformed to the time domain. We denote the resulting matrix by
	\begin{equation}
		\tilde{\mathbf{Y}}_{\mathrm{td}}
		=
		\mathtt{IFFT}\!\left(\widetilde{\mathbf{Y}}_{\mathrm{pilot}}\right)
		=
		\mathtt{IFFT}\!\left(\mathbf{Y}_{\mathrm{pilot}}/\sqrt{P_T}\right).
	\end{equation}
	The matrix $\tilde{\mathbf{Y}}_{\mathrm{td}}\in\mathbb{C}^{N\times M}$ preserves the $N$ antenna channels across $M$ transformed measurement samples. Before extracting the parameters for path $p$, the corresponding residual to that path is calculated as:
	\begin{equation}
		\mathbf{R}_{p-1}[:,m]
		=
		\tilde{\mathbf{Y}}_{\mathrm{td}}[:,m]
		-
		\sum_{l=1}^{p-1}
		\alpha_l\,
		\mathbf{a}(\phi_l,\theta_l)\,
		\mathbf{s}_{\mathrm{td}}(m - \tau_l f_s).
		\label{eq:pathResidue}
	\end{equation}
	
	A first candidate delay estimate is obtained by computing the cross-correlation between the residual and the pilot signal:
	\begin{equation}
		\mathbf{G}
		=
		\mathtt{IFFT}\!\left[
		\mathtt{FFT}(\mathbf{R}_{p-1})
		\odot
		\mathtt{FFT}(\mathbf{s})^{H}
		\right].
	\end{equation}
	
	To define the coarse grid search, let $\mathcal{A}$ denote the coarse azimuth–elevation grid, let $\tau_{\max}$ be the configured delay cap in samples, and let $D_{\mathrm{coarse}}=\min(\tau_{\max},M)$ denote the usable integer-lag search span. The coarse delay grid is $\mathcal{T} = \{0,\Delta_\tau,2\Delta_\tau,\ldots,(D_{\mathrm{coarse}}-1)\Delta_\tau\}$, $\Delta_\tau = 1/f_s$, where $\Delta_\tau$ is the sampling period. The fractional-delay NM refinement then searches the continuous interval $[0,\,D_{\mathrm{coarse}}/f_s)$, while the NM refinement in the angular domain searches the full azimuth–elevation range.
	
	For each integer delay bin $d \in \mathcal{T}$, the initializer evaluates the best beamformed response over the angle grid:
	\begin{equation}
		\Gamma(d)
		=
		\max_{(\theta,\phi)\in \mathcal{A}}
		\left\|
		\mathbf{a}^{H}(\theta, \phi)\,
		\mathbf{G}[:,d]
		\right\|.
	\end{equation}
	Up to eight delay candidates separated by more than two samples are retained. For each retained candidate \(c\), the algorithm refines the delay by solving
	\begin{equation}
		\tau_c^{*}
		=
		\arg\max_{\tau}
		\left\|
		\mathbf{a}^{H}(\phi_c, \theta_c)\,
		\mathbf{z}(\tau)
		\right\|,
		\label{eq:init_tau}
	\end{equation}
	where \((\phi_c, \theta_c)\) is the AoA corresponding to the \(c\)th retained candidate and
	\begin{equation}
		\mathbf{z}(\tau)
		=
		\mathbf{R}_{p-1}\mathbf{s}^H_{\mathrm{td}}(\tau f_s).
	\end{equation}
	
	Given \(\tau_c^{*}\) and \(\mathbf{z}(\tau_c^{*})\), the AoA is updated once more by solving
	\begin{equation}
		(\phi^{*}, \theta^{*})
		=
		\arg\max_{(\phi,\theta)}
		\left\|
		\mathbf{a}^{H}(\phi, \theta)\,
		\mathbf{z}(\tau_c^{*})
		\right\|.
		\label{eq:init_AoAmax}
	\end{equation}
	
	The updates in~\eqref{eq:init_tau} and~\eqref{eq:init_AoAmax} alternate for three sweeps. Then we employ NM algorithm once to refine the scalar delay in~\eqref{eq:init_tau} and the two angular coordinates in~\eqref{eq:init_AoAmax}, producing continuous-valued estimates from the coarse-grid seeds.
	
	Given the refined delay and AoA of candidate $p$, its complex gain is calculated by least-squares (LS) projection:
	\begin{equation}
		\alpha_p = \frac{\mathbf{a}^H(\theta_p^*,\phi_p^*)\mathbf{R}_{p-1}{\mathbf{s}^H_{\mathrm{td}}(\tau_p^* f_s)}}{\max\!\left(\norm{\mathbf{a}(\theta_p^*,\phi_p^*)}_2^2\norm{\mathbf{s}_{\mathrm{td}}}_2^2,\epsilon_{\mathrm{init}}\right)}. \label{eq:closedformA}
	\end{equation}
	Note that in~\ref{eq:closedformA}, $\mathbf{R}\in\mathbb{C}^{N\times M}$, $\mathbf{a}\in\mathbb{C}^{N\times 1}$, and 	$\mathbf{s}\in\mathbb{C}^{M\times 1}$, so the result is a scalar.
	
	Also note that the initializer is executed once per sample outside the differentiable graph and returns a fixed \(P_{\max}\)-candidate state during training and a \(P_{\mathrm{est}}\)-candidate state during inference. The PCN does not control how many candidates are initialized during training; instead, it uses their initial gain ranking to produce differentiable gates. During inference, the smaller \(P_{\mathrm{est}}\) reduces the complexity of the unfolded refinement and reconstruction stages.
	
	In the next section, we discuss the proposed unfolded refiner network, which completes the proposed architecture by enabling fully differentiable path-parameter estimation and channel-complexity recovery.
	
	\subsection{Deep-Unfolded Refiner Network}\label{sec:LMRefiner}
	The refiner is inspired by the classical RIMAX~\cite{rimax} LM procedure and jointly updates the geometric parameters of all \(P_{\max}\) candidates during training and \(P_{\mathrm{est}}\) candidates during inference. Each layer computes ridge-regularized~\cite{ridge} joint gain estimates from the unweighted candidate basis and applies the gates during signal reconstruction, thereby reducing the computational complexity of the refiner stage. The following subsections describe the individual stages of the refiner.
	
	\subsubsection{Geometric State}
	Following the RIMAX spatial-frequency parameterization, the refiner operates on
	\begin{equation}
		\mathbf{x} = [\mathbf{u}_x^T, \mathbf{u}_y^T, \boldsymbol{\tau}_{\mathrm{samp}}^T]^T,
	\end{equation}
	which reparameterizes the physical azimuth, elevation, and delay of the fixed candidate set. The initial state $\mathbf{x}_0$ is supplied by the initializer in Section~\ref{sec:initializer}.
	
	For path $p$, the scaled spatial-frequency coordinates are defined as $ u_{x,p} = 2\pi\frac{d_x}{\lambda}\cos(\theta_p)\cos(\phi_p)$, $u_{y,p} = 2\pi\frac{d_y}{\lambda}\cos(\theta_p)\sin(\phi_p)$, and $		\tau_{\mathrm{samp},p} = \tau_p f_s$, where $d_x$ and $d_y$ are the inter-element spacings, $\lambda$ is the carrier wavelength, and $f_s$ is the sampling frequency.
	
	\subsubsection{Ridge-Regularized Joint Gain Estimation}
	For a given geometric state, the joint ridge-regularized gain estimate is
	computed using the unweighted candidate basis:
	\begin{equation}
		\boldsymbol{\alpha}_{\mathrm{ridge}}(\mathbf{x})
		=
		(\mathbf{G} + d_{\mathrm{ridge}} \mathbf{I}_{P_{\max}})^{-1} \mathbf{m},\label{eq:alpha_ridge}
	\end{equation}
	where $\mathbf{G}$ is the spatio-temporal Gram matrix, expressed as the Hadamard product of the spatial and temporal Gram matrices:
	\begin{align}
		\mathbf{G} &= \mathbf{G}^s\odot \mathbf{G}^t,\label{eq:gram}\\
		\mathbf{G}^s_{p,q} &=  \mathbf{a}^H_p \mathbf{a}_q,\\
		\mathbf{G}^t_{p,q} &=  \mathbf{s}^H_{p}\mathbf{s}_{q}.	
	\end{align}
	Note that since $\mathbf{G}$ is a rank-1 matrix, the inversion in~\eqref{eq:alpha_ridge} can be computed efficiently using the matrix inversion lemma. For the current geometric state, each candidate path \(p\) is represented through its steering vector and delayed sounding sequence,
	\begin{align}
		\mathbf{a}_p &= \mathbf{a}(u_{x,p}, u_{y,p}) \in \mathbb{C}^{N},\\
		\mathbf{s}_p &= \mathbf{s}[\tau_{\mathrm{samp},p}] \in \mathbb{C}^{M},
	\end{align}
	which together define the matched-filter right-hand side,
	\begin{align}
		\mathbf{m}_p &= \mathbf{a}_p^{H}\mathbf{z}_{p},\\
		\mathbf{z}_{p} &= \mathbf{Y}_{\mathrm{td}}\,\overline{\mathbf{s}_p}.
	\end{align}
	To stabilize the subsequent gain fit, a trainable raw ridge parameter \(\beta_{\alpha}^{\mathrm{raw}}\) produces the regularization term
	\begin{align}
		d_{\mathrm{ridge}}
		&= \max\!\left(\mathrm{softplus}(\beta_{\alpha}^{\mathrm{raw}}) P_s,
		d_{\mathrm{floor}}\right),\\
		d_{\mathrm{floor}}
		&= \epsilon_{\mathrm{ridge}}\times\max\!\left(P_s,\,
		(\overline{G} + 1)\right),
	\end{align}
	where \(P_s=\|\mathbf{s}\|_2^2\) is the sounding-sequence energy, \(\overline{G}=\tfrac{1}{P_{\max}}\mathtt{tr}(\mathbf{G})\), \(\epsilon_{\mathrm{ridge}}=10^{-6}\), and \(\mathrm{softplus}(x)=\log(1+e^{x})\).
	This regularization is applied when computing the joint ridge-regularized gain estimate \(\boldsymbol{\alpha}_{\mathrm{ridge}}\) from the full, unweighted candidate basis. Because this gain-solve is performed before any gate weighting is applied, \(\boldsymbol{\alpha}_{\mathrm{ridge}}\) does not minimize the gated LS objective. Instead, it provides a numerically stable intermediate estimate that the unfolded refinement layers subsequently improve.

	\subsubsection{Gated Residual and Jacobian}
	
	The LM step is obtained by linearizing the residual of each path with respect to the geometric parameters, which requires the Jacobian of that residual. Accordingly, we first form the residual and then compute its Jacobian. With soft gates included, the complex residual is	
	\begin{equation}
		\mathbf{r}_p(\mathbf{x})
		=
		\mathbf{Y}_{\mathrm{td}}
		-
		\sum_{l=1\ l\neq p}^{P_{\max}}
		\mathbf{g}_l\,\alpha_l(\mathbf{x})\,
		\mathbf{a}_l(\mathbf{x})\,
		\mathbf{s}_l^{\mathsf T}(\mathbf{x}),
	\end{equation}
	where $\alpha_p(\mathbf{x})=[\boldsymbol{\alpha}_{\mathrm{ridge}}(\mathbf{x})]_p$
	is the geometry-dependent gain from the ungated ridge fit.
	
	At LM layer $\ell$, the refiner works with the real-valued residual
	\begin{equation}
		\mathbf{r}_{\ell}
		=
		\begin{bmatrix}
			\Re\!\left(\mathbf{r}(\mathbf{x}_{\ell})\right) \\
			\Im\!\left(\mathbf{r}(\mathbf{x}_{\ell})\right)
		\end{bmatrix}
		\in \mathbb{R}^{2MN},
	\end{equation}
	where vectorization is implicit and the real representation enables a
	real-valued LM system. Therefore, the corresponding Jacobian is
	\begin{equation}
		\mathbf{J}_{\ell}
		=
		\frac{\partial \mathbf{r}_{\ell}}{\partial \mathbf{x}_{\ell}}
		\in \mathbb{R}^{2MN \times 3P_{\max}}.
	\end{equation}
	
	Because the gain estimates $\alpha_p(\mathbf{x})$ depend on the geometric state through the ridge-regularized least-squares solve, automatic differentiation propagates through both the gain computation and the gated synthesis. As a result, the Jacobian captures the sensitivity of the gains as well as the residual. Importantly, it is not the profiled Jacobian associated with variable-projection methods, as used in RIMAX, because
	$\boldsymbol{\alpha}_{\mathrm{ridge}}(\mathbf{x})$ is obtained from an ungated ridge-regularized solve rather than as the conditional minimizer of the gated LS objective:
	
	\begin{equation}
		\mathbf{J}_{\ell}
		=
		\frac{\partial}{\partial \mathbf{x}_{\ell}}
		\mathbf{r}\!\left(\mathbf{x}_{\ell},\,
		\boldsymbol{\alpha}_{\mathrm{ridge}}(\mathbf{x}_{\ell})\right).
	\end{equation}
	This Jacobian is obtained directly via automatic differentiation.
	
	\subsubsection{Gauss-Newton Matrix and Score}
	
	Given the gain-coupled residual Jacobian at layer $\ell$, the Gauss-Newton curvature approximation and residual score are
	\begin{equation}
		\mathbf{F}_{\ell} =\mathbf{J}_{\ell}^{\mathrm{T}}\mathbf{J}_{\ell} \text{, and }
		\mathbf{S}_{\ell}=	\mathbf{J}_{\ell}^{\mathrm{T}}	\mathbf{r}_{\ell}.
	\end{equation}
	Here the matrix $\mathbf{F}_{\ell}$ is positive semidefinite and approximates the local curvature of the LS objective, while $\mathbf{S}_{\ell}$ provides the corresponding score. We retain the symbol $\mathbf{F}_{\ell}$ because this matrix is like Fisher-information (FIM) under the adopted Gaussian residual model.	
	
	\subsubsection{Learned Damping and Positive-Definite Floor}
	
	Each LM layer of the refiner network introduces trainable parameters that control the stability and effectiveness of the LM update. These parameters enter the \emph{LM system matrix}, which is constructed from the FIM and governs the curvature of the GN step. We begin by describing the learned damping parameters and their role.
	
	In classical LM, the damping parameter is adjusted adaptively through an accept/reject loop~\cite{rimax}, which increases algorithmic complexity and introduces non-differentiable behavior. In the proposed deep-unfolded LM refiner, this mechanism is replaced by trainable, layer-wise damping parameters\ie $\boldsymbol{\lambda}_{\ell}^{\mathrm{raw}} \in \mathbb{R}^{3}$. To ensure positivity, these raw parameters are passed through a softplus activation\ie $\boldsymbol{\lambda}_{\ell} = \mathrm{softplus}\!\left(\boldsymbol{\lambda}_{\ell}^{\mathrm{raw}}\right) \in \mathbb{R}_{>0}^{3}$.
	
	Because $(u_x,u_y,\tau_{\mathrm{samp}})$ have different scales and curvature, each layer uses one damping value per parameter type. During training, these values are expanded to length $3P_{\max}$:
	\begin{equation}
		\boldsymbol{\lambda}_{\ell}^{(P_{\max})}=
		\begin{bmatrix}
			\lambda_{\ell,x}\, \mathbf{1}_{P_{\max}} \\
			\lambda_{\ell,y}\, \mathbf{1}_{P_{\max}} \\
			\lambda_{\ell,\tau}\, \mathbf{1}_{P_{\max}}
		\end{bmatrix},
	\end{equation}
	where \(\mathbf{1}_n\) denotes a vector of ones of length \(n\). These damping values are shared across all candidate paths within a layer, ensuring consistent regularization for each parameter type while keeping the LM system well-scaled.
	
	To further enhance numerical stability, especially during training, and to prevent near-singularity of the LM system matrix, a Fisher-information-relative floor is added:
	\begin{equation}
		\mu_{\ell} 	=\lambda_{\mathrm{floor.rel}}\cdot \max\!\left(\frac{1}{3P_{\max}}\mathrm{tr}(\mathbf{F}_{\ell}),	10^{-12}\right).
	\end{equation}
	This term scales with the average GN curvature and improves conditioning of the LM system.
	
	Combining all components, the LM system matrix at layer $\ell$ is described as:
	\begin{equation}
		\mathbf{A}_{\ell}=\mathbf{F}_{\ell}	+\mathrm{diag}\left(\boldsymbol{\lambda}_{\ell}^{(P_{\max})}\right)	+\mu_{\ell}\mathbf{I}_{3P_{\max}}.
	\end{equation}
	The resulting system replaces online cost-based damping adaptation of classical LM with a fixed, trainable layer-wise schedule.
	
	\subsubsection{LM Update}
	Given the LM system matrix $\mathbf{A}_{\ell}$ and the score vector $\mathbf{S}_{\ell}$, the LM update direction at layer $\ell$ is obtained by solving the linear system
	\begin{equation}
		\mathbf{A}_{\ell}\, \boldsymbol{\delta}_{\ell} 	=	\mathbf{S}_{\ell},
	\end{equation}
	where $\boldsymbol{\delta}_{\ell}\in\mathbb{R}^{3P_{\max}}$ is the ridge-regularized GN direction used to update the geometric parameters at that layer.
	
	Each LM layer also includes a trainable step-trust scalar that controls the magnitude of the update:
	\begin{equation}
		\eta_{\ell} =\sigma\left(\eta_{\ell}^{\mathrm{raw}}\right) \in (0,1),
	\end{equation}
	where $\sigma(\cdot)$ is the sigmoid function. The scalar is learned for each layer and is shared across samples and candidate paths.
	
	The geometric state is then updated according to
	\begin{equation}
		\mathbf{x}_{\ell+1} =\mathbf{x}_{\ell}-	\eta_{\ell}\, \boldsymbol{\delta}_{\ell},
	\end{equation}
	
	\subsubsection{Deep Supervision}
	
	At each layer $\ell$, the current geometric state $(\mathbf{u}_{x,\ell}, \mathbf{u}_{y,\ell}, \boldsymbol{\tau}_{\mathrm{samp},\ell})$
	and the ridge-regularized gain estimates $\boldsymbol{\alpha}_{\mathrm{ridge},\ell}$ are used to synthesize a reconstruction of the received pilot matrix using~\eqref{eq:systemModel} and~\eqref{eq:channelmodelGate}\ie
	\begin{equation}
		\hat{\mathbf{Y}}_{\ell} = \mathrm{Synthesise} \big(\mathbf{u}_{x,\ell},\mathbf{u}_{y,\ell},\tau_{\mathrm{samp},\ell}, \boldsymbol{\alpha}_{\mathrm{ridge},\ell},	\mathbf{g}	\big).
	\end{equation}
	where $\mathbf{g}$ is the gate vector defined in Section~\ref{sec:PCN}. The gates weight all candidate contributions continuously during training.
	
	Each reconstruction is then compared to the measured time domain signal in the pilot $\mathbf{Y}_{\mathrm{td}}$, and the resulting discrepancy contributes to the LM supervision loss defined as:
	\begin{equation}
		\mathcal{L}_{\mathrm{LM}} =	\frac{1}{L}	\sum_{\ell=1}^{L}\frac{
			\left\| \mathbf{Y}_{\mathrm{td}} - \hat{\mathbf{Y}}_{\ell}\right\|_{F}^{2}}{	\max\!\left(\left\|\mathbf{Y}_{\mathrm{td}}\right\|_{F}^{2},\epsilon_{\mathrm{tiny}}	\right)}.\label{eq:loss_DS}
	\end{equation}
	Here, $\hat{\mathbf{Y}}_{\ell}$ denotes the reconstructed received signal
	obtained from the geometric state and ridge-regularized gain estimates at layer $\ell$. This term provides a reconstruction objective at every unfolded depth and directly trains the damping and step-trust schedules.
	
	\subsection{Training Loss}\label{sec:trainingLoss}
	
	The training objective combines final reconstruction accuracy, layer-wise LM deep supervision, and path-count parsimony.
	
	\subsubsection{Final Reconstruction NMSE}
	The normalized mean-squared error (NMSE) between the measured and reconstructed time-domain pilot is given by
	\begin{equation}
		\mathcal{L}_{\mathrm{ch}}=\frac{\norm{\mathbf{Y} - \hat{\mathbf{Y}}}_{F}^{2}}{		\max\!\left(\norm{\mathbf{Y}}_{F}^{2},\epsilon_{\mathrm{tiny}}\right)},
	\end{equation}
	where $\hat{\mathbf{Y}}$ is the reconstructed received signal obtained from the
	final estimated path parameters, and $\epsilon_{\mathrm{tiny}}$ is a small numerical stability constant.
	
	\subsubsection{Path-Count Parsimony}
	To encourage sparse path selection, a path-count penalty is included:
	\begin{equation}
		\mathcal{L}_{\mathrm{count}} =\frac{P_{\mathrm{est}}}{P_{\max}}.
	\end{equation}
	
	\subsubsection{Complete Training Objective}
	The reconstruction NMSE, the deep-supervision loss defined in
	\eqref{eq:loss_DS}, and the path-count parsimony together form the training
	objective:
	\begin{equation}
		\mathcal{L} =\mathcal{L}_{\mathrm{ch}}+	\lambda_{\mathrm{LM}}\,\mathcal{L}_{\mathrm{LM}}+ \lambda_{\mathrm{count}}\,\mathcal{L}_{\mathrm{count}},
	\end{equation}
	where $\lambda_{\mathrm{LM}}$ and $\lambda_{\mathrm{count}}$ are regularization
	weights that balance the contributions of the different loss components.

	\subsection{Training, Validation, and Test Data Acquisition}
	
	MPSi requires the received signal matrix in the pilot phase $\mathbf{Y}_{\mathrm{td}}$, the known sounding sequence, and the pilot transmit power. In a practical deployment, these observations could be collected at different user locations throughout a serving cell.
	
	Because measured BS data were unavailable for this study, we relied on simulated channels to evaluate the proposed method. We used the Sionna ray tracer~\cite{sionna} with 3D environments constructed from OpenStreetMap data~\cite{openstreetmap}; user locations were uniformly sampled across each environment, and the BS position and height were selected based on publicly available operator deployment information.
	
	For each user position, the ray tracer produced the geometric path parameters (AoA, delays, and path gains). These parameters were then used to synthesize the channel and generate the corresponding received pilot matrices according to the system model in~\eqref{eq:systemModel}, with additive noise applied as described in Section~\ref{sec:systemModel}. This process effectively emulates placing users throughout the environment and measuring the resulting channels. 
	
	Importantly, the ray-traced path parameters are not used as training labels. Training relies solely on the received pilot measurements, the known sounding sequence, and the transmit power through the loss function defined in Section~\ref{sec:trainingLoss}. The role of the ray tracer is only to generate the pilot observations used in this study. Consequently, while Sionna is used here to create realistic training and test datasets, the proposed MPSi framework can be trained using real measurement data whenever received pilot observations are available.
	
	\subsection{A Note on the Complexity of MPSi}
	\begin{figure}[t]
		\begin{subfigure}[b]{0.49\linewidth}
			\centering
			\includegraphics[width=0.75\textwidth]{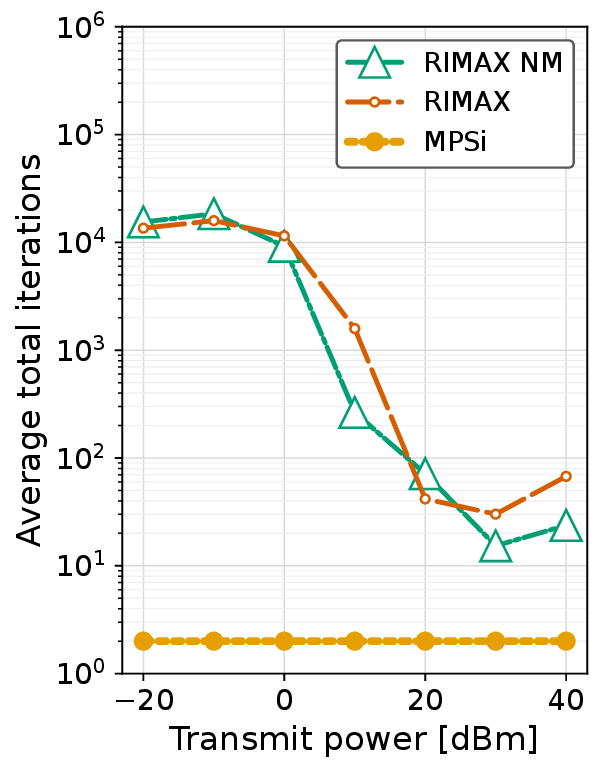}
			\caption{}
			\label{fig:iter_bahen}
		\end{subfigure}
		\hfill
		\begin{subfigure}[b]{0.49\linewidth}
			\centering
			\includegraphics[width=0.75\textwidth]{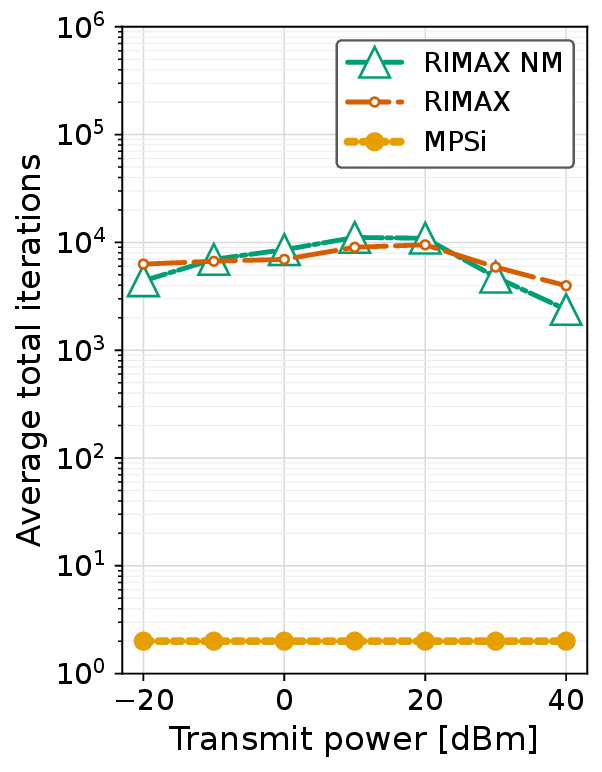}
			\caption{}
			\label{fig:iter_dundas}
		\end{subfigure}
		\caption{The number of iterations in proposed MPSi compared to the RIMAX algorithm in Bahen and Dundas scenarios, respectively.}
		\label{fig:iterations}
	\end{figure}
	
	The proposed MPSi algorithm provides low-complexity yet accurate estimation of communication-path parameters. To contextualize its complexity, we compare MPSi with classical RIMAX, an accurate but computationally heavier algorithm designed for related geometric channel-estimation tasks.
	
	Classical RIMAX performs data-dependent model-order updates together with cost-checked LM refinement~\cite{rimax}, resulting in a variable number of iterations across observations. Its prune-and–add strategy keeps the LM complexity in check: the number of active paths is initialized conservatively, and at each outer iteration the stability of the current paths is evaluated. Paths are then added or removed until convergence, causing the iteration count to depend on the information content of the received signal.
	
	By contrast, at inference, the estimated number of active paths \(P_{\mathrm{est}}\) in MPSi directly reduces the LM system to a \(3P_{\mathrm{est}} \times 3P_{\mathrm{est}}\) matrix, lowering complexity relative to the training phase. Figure~\ref{fig:iterations} compares the average number of iterations used by both methods: RIMAX reports the sum of its inner iterations across all outer iterations, whereas MPSi always performs a fixed number \(L_{\mathrm{LM}} = 2\) of unfolded layers used in the experiments. For this comparison, RIMAX-NM uses the initialization method described in Section~\ref{sec:initializer}, while classical RIMAX uses its standard initialization without NM.
	
	As depicted in Figure~\ref{fig:iterations}, the number of RIMAX iterations depends on the information content of the received signal: higher transmit power yields stronger resolvable paths and therefore fewer iterations. MPSi, by construction, executes a fixed number of unfolded layers at every pilot power, independent of \(P_T\). This comparison shows that MPSi achieves a bounded and substantially smaller refinement depth than RIMAX. it is worth noting that iteration count does not represent algorithmic latency; in this paper it is used solely as a control-flow complexity indicator. Deployment-level latency claims require operation counts and timing
	measurements on a specified hardware and software platform with optimized implementations of both algorithms~\cite{executiontime}, which is beyond the scope of this paper.
	
	\section{Simulation Results}\label{sec:simulations}
	This section evaluates MPSi for initial access, full-channel estimation, and pseudo-label generation. The BS employs a $32$-element UPA with eight horizontal and four vertical elements, half-wavelength spacing, a carrier frequency of $5$~GHz, and a nominal baseband measurement bandwidth of $100$~MHz.
	
\begin{figure}[t]
		\begin{subfigure}[b]{0.49\linewidth}
			\centering
			\includegraphics[width=0.97\textwidth]{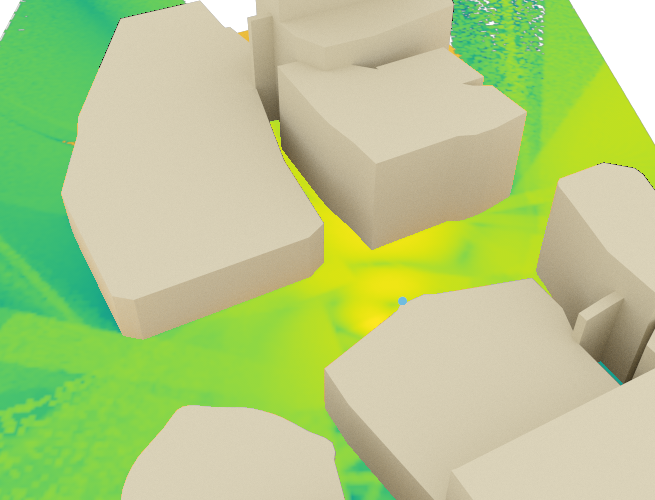}
			\caption{}
			\label{fig:cm_bahen}
		\end{subfigure}
		\hfill
		\begin{subfigure}[b]{0.49\linewidth}
			\centering
			\includegraphics[width=0.97\textwidth]{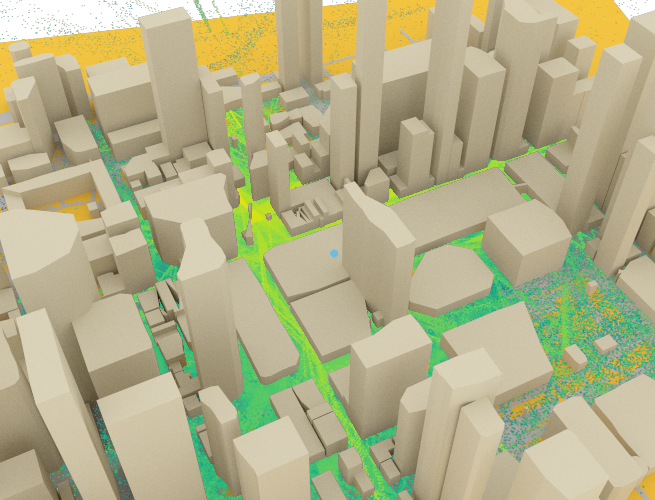}
			\caption{}
			\label{fig:cm_dundas}
		\end{subfigure}
		
		\caption{Analyzed maps and corresponding BS coverage zones. The BS locations are shown with blue dots: (a) the University of Toronto campus (Bahen Centre), and (b) the Yonge–Dundas intersection in downtown Toronto, ON, Canada.}
		\label{fig:cm}
	\end{figure}

	For each realization, a user is placed at a sampled map location, and Sionna RT~\cite{sionna} generates the propagation paths for the configured scene. These parameters are used with~\eqref{eq:channelmodel} and~\eqref{eq:systemModel} to synthesize the channel and the corresponding noisy pilot observation. The receiver-noise power is set to $-95$~dBm. We consider two Toronto environments: the Bahen Centre at the University of Toronto and the Yonge-Dundas intersection in the city core, referred to as the Bahen and Dundas scenarios, respectively. These environments, therefore, represent a small cell and a dense urban area, as shown in Fig.~\ref{fig:cm}.  The Bahen and Dundas scenarios use $M=64$ and $M=256$ pilot FFT bins, respectively, with bin spacings $100$~MHz$/M$; full-band channel and BER evaluation uses $1024$ FFT bins over the same bandwidth. MPSi is trained separately for each environment using $512$ realizations, a batch size of $32$, and the Adam optimizer~\cite{adam} with a learning rate of $0.0003$. The test results are obtained from $100$ additional realizations generated independently of the training data. Table~\ref{tbl:params} summarizes the principal learning parameters.
	
	Note that in the following analysis, the empirical cumulative distribution function (CDF) for a metric $x$ evaluated over $R$ realizations is defined as $\widehat{F}_{x}(z) = \frac{1}{R}\sum_{r=1}^{R}\mathbb{1}\left[x_r \leq z\right]$. Here $x_r$ denotes the metric value for realization $r$ and $\mathbb{1}[A]$ is the indicator function, which equals one if its argument is true and zero otherwise.
	
	\begin{table}[t]
		\caption{Simulation Parameters}
		\label{tbl:params}
		\centering
	\scriptsize
		\begin{tabular}{lcc}
			\hline
			\textbf{Parameter} & \textbf{Bahen} & \textbf{Dundas} \\
			\hline
			\noalign{\smallskip}
			Training $P_T$  & $10$~dBm & $30$~dBm \\
			$M$                           & $64$ & $256$ \\
			Configured delay cap $\tau_{\max}$ [samples] & $80$ & $350$ \\
			PCN structure				 & \multicolumn{2}{c}{$M \times 64 \times 32 \times 1$; $M$-sample input} \\
			$P_{\min},\, P_{\max}$       & \multicolumn{2}{c}{$2,\; 20$} \\
			$\ell_{\mathrm{PCN}}$		 & \multicolumn{2}{c}{$2$}\\
			$\lambda_{\mathrm{LM}}$      & \multicolumn{2}{c}{$0.5$} \\
			$\lambda_{\mathrm{count}}$   & \multicolumn{2}{c}{$0.02$} \\
			$\lambda_{\mathrm{floor.rel}}$ & \multicolumn{2}{c}{$0.001$} \\
			$\epsilon_{\mathrm{tiny}}$   & \multicolumn{2}{c}{$1.18\times10^{-38}$ (float32 minimum)} \\
			Total number of subcarriers $N_{scr}$  & \multicolumn{2}{c}{1024} \\
			\hline
		\end{tabular}
	\end{table}

	\subsection{Initial Access Analyses}
	
	To demonstrate the performance of the MPSi in initial access application, we compare its performance with three beam-search baselines representing different accuracy-overhead tradeoffs:
	
	\begin{itemize}
		\item \textbf{DFT codebook (dense-search reference):} This baseline exhaustively searches a fixed $32\times16$ azimuth-elevation grid containing $512$ DFT-style beams. Each beam is projected onto the received pilot, and the beam with the largest projected power is selected. It provides a dense angular-quantization reference at high search overhead.
		
		\item \textbf{Hierarchical codebook (coarse-to-fine search):} This method first evaluates an $8\times4$ coarse grid containing 32 beams. It then allocates the remaining 32 checks from a shared 64-beam budget to local refinement around the strongest coarse direction. The final direction is selected from all beams evaluated across both stages.
		
		\item \textbf{3GPP-inspired SSB sweep:} This baseline searches a fixed codebook of 64 uniformly distributed beams and selects the beam with maximum received pilot power. Each modeled synchronization signal block (SSB) occasion is assigned four OFDM symbols and 240 resource elements.
	\end{itemize}
	
	\begin{figure*}[t]
		\begin{subfigure}[b]{0.49\textwidth}
			\centering
			\includegraphics[width=0.7\textwidth]{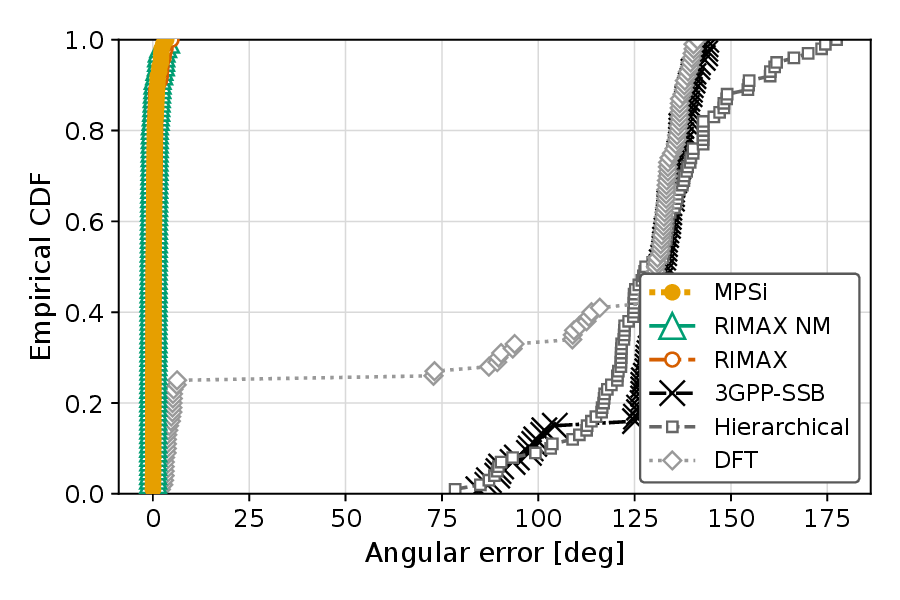}
			\caption{}
			\label{fig:CDF_Bahen_ang}
		\end{subfigure}
		\hfill
		\begin{subfigure}[b]{0.49\textwidth}
			\centering
			\includegraphics[width=0.7\textwidth]{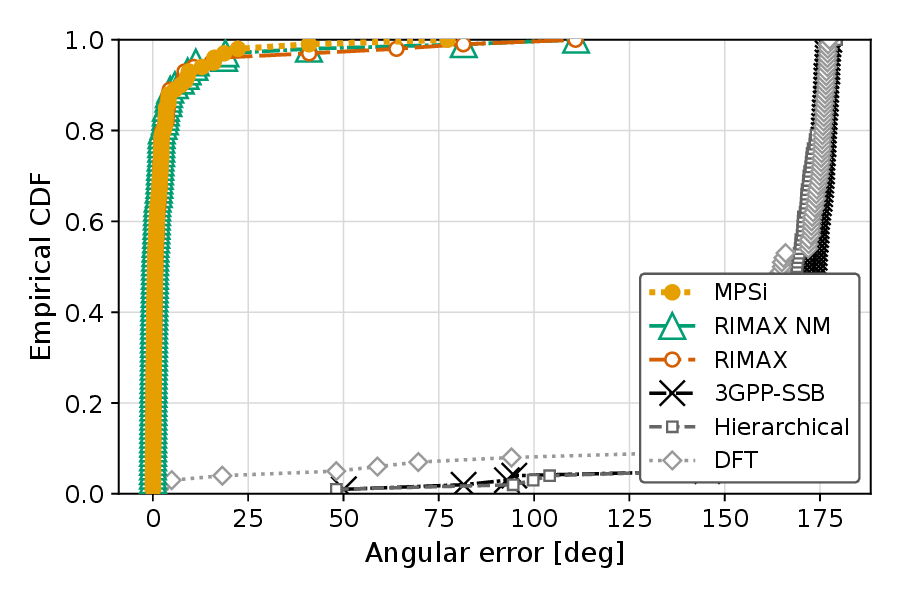}
			\caption{}
			\label{fig:CDF_Dundas_ang}
		\end{subfigure}
		\hfill
		\begin{subfigure}[b]{0.49\textwidth}
			\centering
			\includegraphics[width=0.7\textwidth]{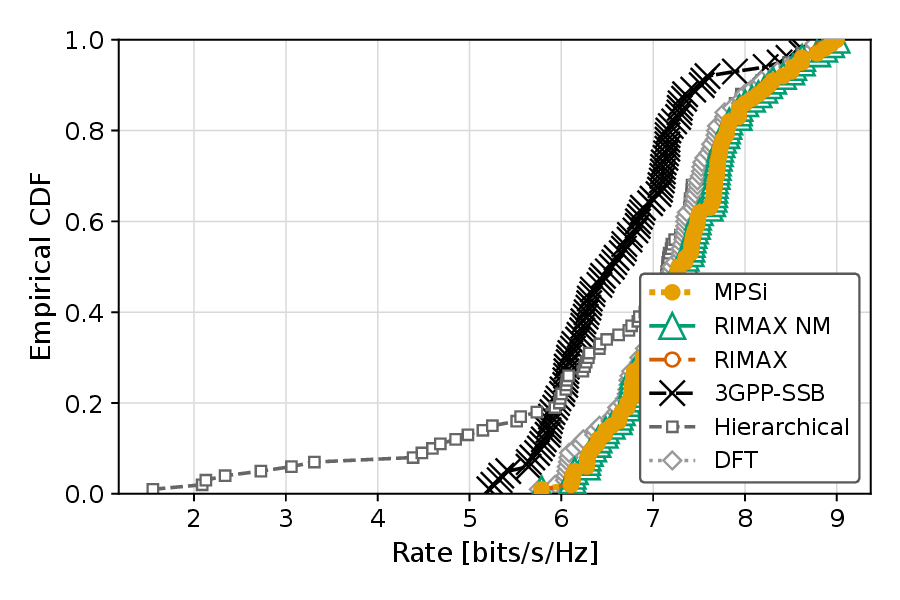}
			\caption{}
			\label{fig:CDF_Bahen_rate}
		\end{subfigure}
		\hfill
		\begin{subfigure}[b]{0.49\textwidth}
			\centering
			\includegraphics[width=0.7\textwidth]{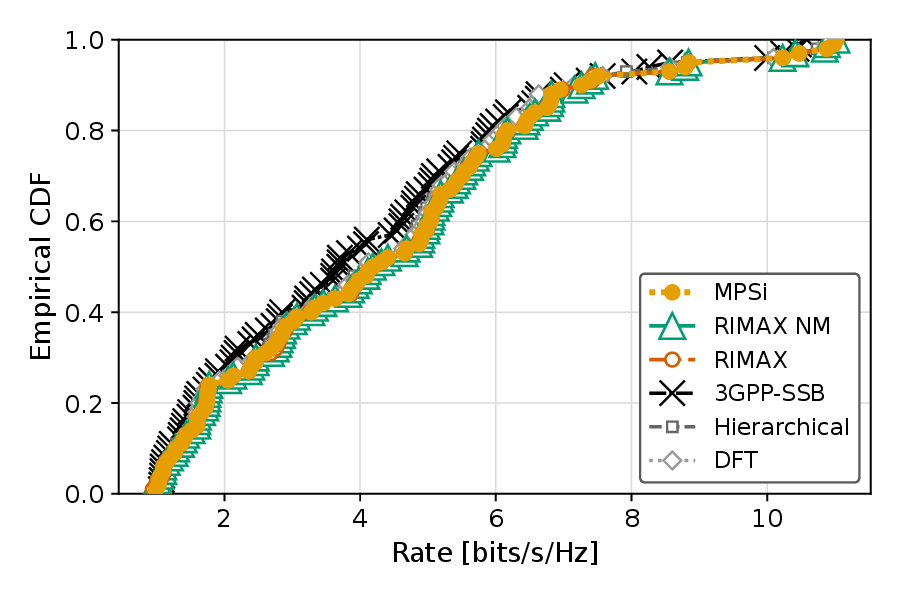}
			\caption{}
			\label{fig:CDF_Dundas_rate}
		\end{subfigure}
		
		\caption{Initial access performance comparison across codebook-based and estimator-based designs when $P_T = -10$~dBm in Bahen and $P_T = 10$~dBm in Dundas scenario. Subfigures (a) and (b) show the CDF of the dominant-path angular error, while (c) and (d) present the CDF of the achievable spectral efficiency for the Bahen and Dundas maps, respectively.}
		\label{fig:CDFs}
	\end{figure*}
	
	Figure~\ref{fig:CDFs} compares estimator-based and codebook-based initial access in the Bahen and Dundas scenarios using empirical cumulative distribution functions (CDFs) of achievable spectral efficiency and dominant-path angular error. 
	
	Estimator-based methods generally achieve lower angular error and higher achievable rates than lower-resolution codebooks because they recover continuous AoAs, whereas codebooks are restricted to predefined beam directions. Among the codebook baselines, the dense DFT codebook attains the closest accuracy to estimator-based methods by using a finer angular grid, but it requires $512$ beam checks. By contrast, the hierarchical and 3GPP-inspired designs operate with much smaller search budgets and therefore exhibit larger quantization errors. Among the estimator-based approaches, MPSi delivers initial-access performance comparable to RIMAX while using a fixed number of unfolded LM layers and thus, as described earlier, lower complexity.
	
	As depicted in Fig.~\ref{fig:CDFs}(d), the achievable rates in the dense urban environment of Dundas are very similar across both codebook and estimator-based methods. This is because the multipath richness in this
	environment is extremely high, so only a small fraction of the received power is carried by the dominant path. The resolution of the codebook search and the array beamforming are therefore sufficient to capture the dominant-path contribution accurately, resulting in comparable achievable rates across all methods.
	
\begin{figure*}[t]
		\begin{subfigure}[b]{0.49\textwidth}
			\centering
			\includegraphics[width=0.7\textwidth]{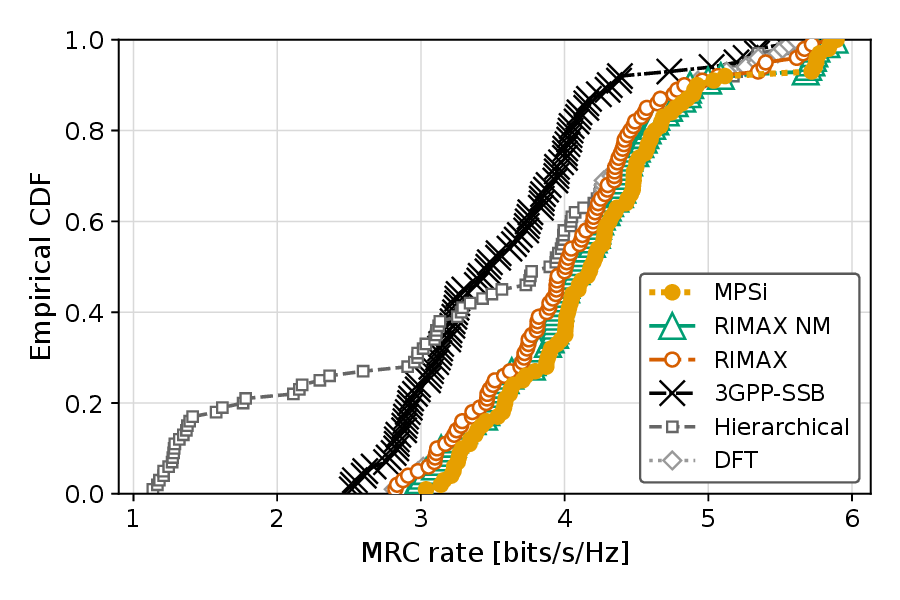}
			\caption{}
			\label{fig:Bahen_CDF_mrcrate}
		\end{subfigure}
		\hfill
		\begin{subfigure}[b]{0.49\textwidth}
			\centering
			\includegraphics[width=0.7\textwidth]{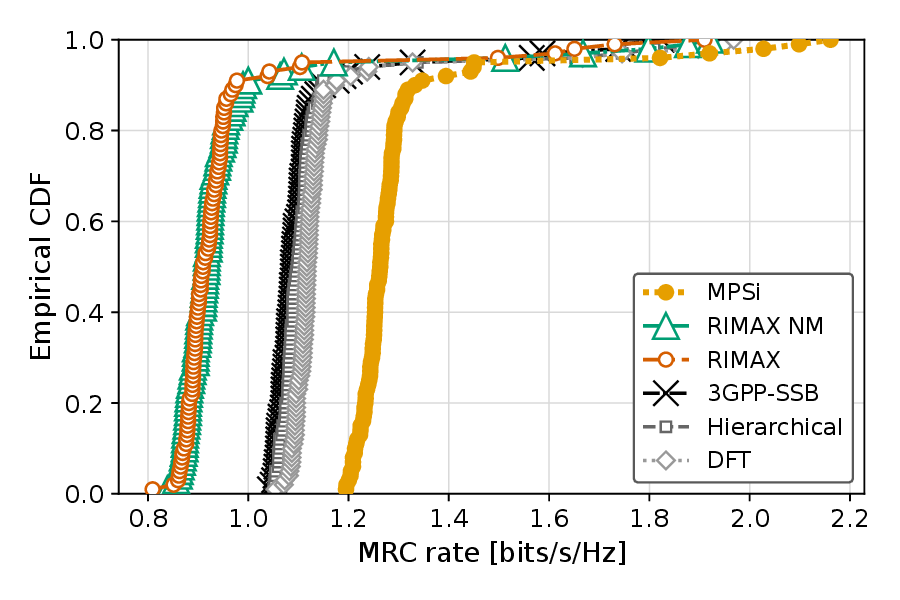}
			\caption{}
			\label{fig:Dundas_CDF_mrcrate}
		\end{subfigure}
		\caption{Empirical CDF of the uplink achievable rate obtained using all communication paths for (a) Bahen and (b) Dundas.}
		\label{fig:CDF_mrcrate}
	\end{figure*}

	A further distinction between the two methods is that estimators recover the parameters of multiple propagation paths, whereas codebook methods typically provide only a selected beam direction or dominant-path AoA. This richer multipath representation can support higher communication rates during initial access, particularly at low pilot powers. As shown in Fig.~\ref{fig:CDF_mrcrate}, with $P_T=-20$~dBm, MPSi achieves a substantially higher data rate than the estimator- and codebook-based baselines. This improvement results from its ability to identify multiple active paths, estimate their parameters, and reconstruct the full communication channel.
	
	\begin{table}[t]
		\centering
		\scriptsize
		\setlength{\tabcolsep}{3pt}
		\caption{Resource usage of different methods in initial access evaluated on Bahen scenario}
		\label{tab:initial_access_resources}
		\begin{tabular}{lrrrrr}
			\toprule
			Method & Time & Freq/tx & Pilot length & \# of Symbols & Air Latency [ms] \\
			\midrule
			MPSi 		& 1 & 64 & 64 & 2 & 0.018 \\
			RIMAX NM 	& 1 & 64 & 64 & 2 & 0.018 \\
			RIMAX 		& 1 & 64 & 64 & 2 & 0.018 \\
			3GPP-SSB 	& 64 & 240 & 15360 & 320 & 2.858 \\
			Hierarchical & 64 & 64 & 4096 & 128 & 1.143 \\
			DFT 		& 512 & 64 & 32768 & 1024 & 9.144 \\
			\bottomrule
		\end{tabular}
	\end{table}
	
	To compare the initial-access resource requirements, Table~\ref{tab:initial_access_resources} summarizes the time-frequency usage and modeled air-interface latency for the Bahen scenario. The ``Time'' column denotes the number of pilot transmissions or beam occasions, while ``Freq/tx'' gives the number of frequency-domain resource elements per transmission. Their product gives the total pilot-resource count. The estimator-based methods use a single transmission over 64 frequency elements, corresponding to 64 pilot elements and two modeled air-interface symbols when one sounding and one feedback symbol are counted. Each air-interface latency is set to $0.018$~ms. 
	
	The codebook-based methods require substantially more pilot transmissions because they evaluate multiple beam directions to select the strongest one. By contrast, the estimator-based methods obtain the received signal over multiple subcarriers in a single transmission and estimate the dominant-path AoA from that observation, albeit with higher computational complexity. Note that the resource accounting in Table~\ref{tab:initial_access_resources} represents air-interface overhead and not the computational complexity. A codebook check primarily involves beam projection or matched filtering, whereas the estimator-based methods require nonlinear multipath estimation resulting in much higher computational complexity.
	
	\subsection{Full Multipath-Channel Estimation Performance}
	
	Beyond dominant-path estimation for initial access, MPSi reconstructs the full modeled multipath channel. We evaluate the reconstruction over all $1024$ subcarriers using channel normalized mean-square error (NMSE) and bit-error rate (BER). We compare MPSi with RIMAX variants, perfect channel knowledge, and least-squares (LS) channel estimation.
	
\begin{figure*}[t]
		\begin{subfigure}[b]{0.49\textwidth}
			\centering
			\includegraphics[width=0.7\textwidth]{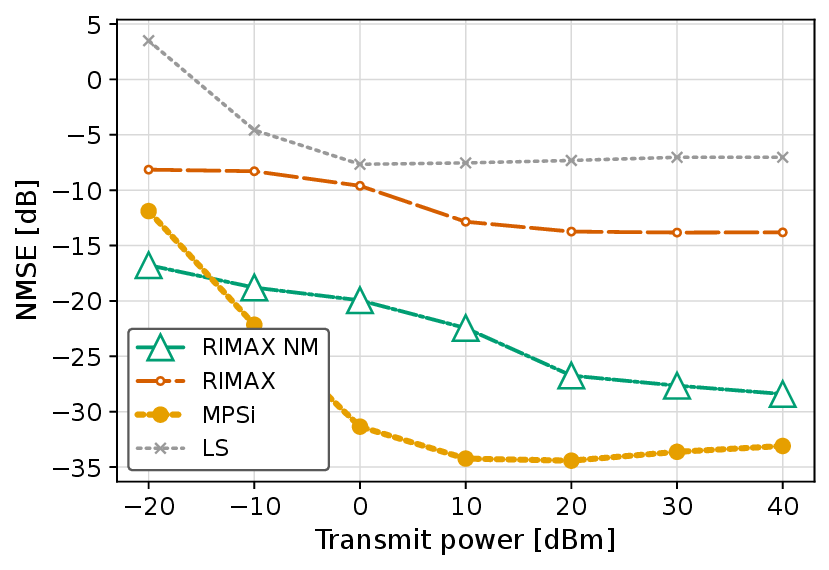}
			\caption{}
			\label{fig:Bahen_NMSE}
		\end{subfigure}
		\hfill
		\begin{subfigure}[b]{0.49\textwidth}
			\centering
			\includegraphics[width=0.7\textwidth]{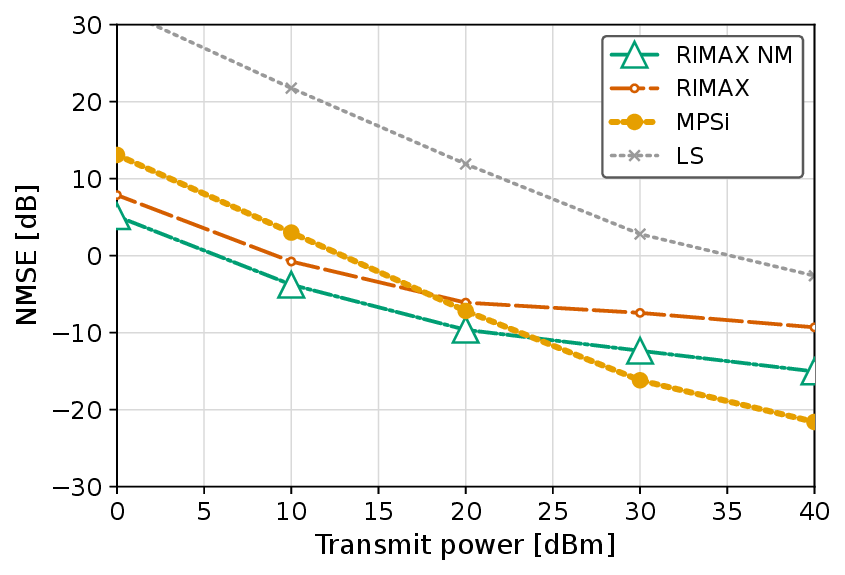}
			\caption{}
			\label{fig:Dundas_NMSE}
		\end{subfigure}
		\caption{Full-band channel NMSE versus pilot transmit power for (a) Bahen and (b) Dundas.}
	\label{fig:NMSE_performance}
	\end{figure*}

	Figure~\ref{fig:NMSE_performance} shows that MPSi and RIMAX with NM initialization achieve low channel NMSE across the evaluated subcarriers. In the Bahen and Dundas scenarios, the NMSE reaches approximately $-30$~dB and $-20$~dB, respectively, demonstrating the accuracy of the proposed estimator. These results indicate that the estimated path parameters are sufficiently accurate to reconstruct the frequency-dependent channel. Delay estimates are particularly important for frequency extrapolation because each path's phase varies with subcarrier frequency, while accurate AoA estimates support reconstruction of the spatial channel response across the BS array.
	
	MPSi also outperforms RIMAX with NM initialization, partly because MPSi treats scattering components as resolvable paths that contribute to communication performance, whereas RIMAX models them as noise. Treating these components as noise introduces additional channel-modeling error within the evaluated fading window. By contrast, LS estimation and RIMAX without NM initialization exhibit higher channel NMSE. The LS baseline does not model parametric path delays and is therefore less suited to reconstructing unmeasured frequencies. The difference between the RIMAX variants further illustrates the sensitivity of nonlinear path refinement to initialization.
	
	We further evaluate the practical impact of channel-estimation accuracy through the bit-error rate (BER) achieved in a single-user uplink scenario with maximum-ratio combining (MRC). Equiprobable bits are Gray mapped to the selected 64-quadrature amplitude modulation (64-QAM) constellation and detected using each channel estimate.
	
\begin{figure*}[t]
		\begin{subfigure}[b]{0.49\textwidth}
			\centering
			\includegraphics[width=0.7\textwidth]{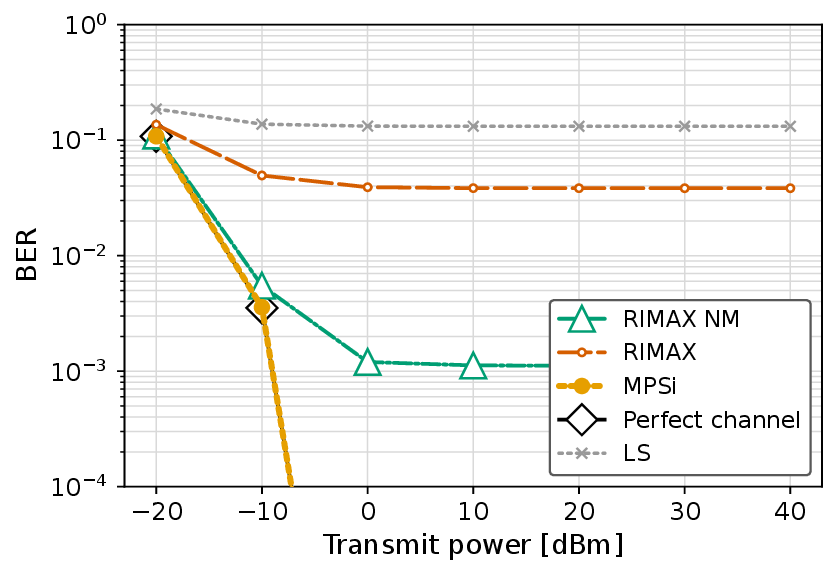}
			\caption{}
			\label{fig:Bahen_ber}
		\end{subfigure}
		\hfill
		\begin{subfigure}[b]{0.49\textwidth}
			\centering
			\includegraphics[width=0.7\textwidth]{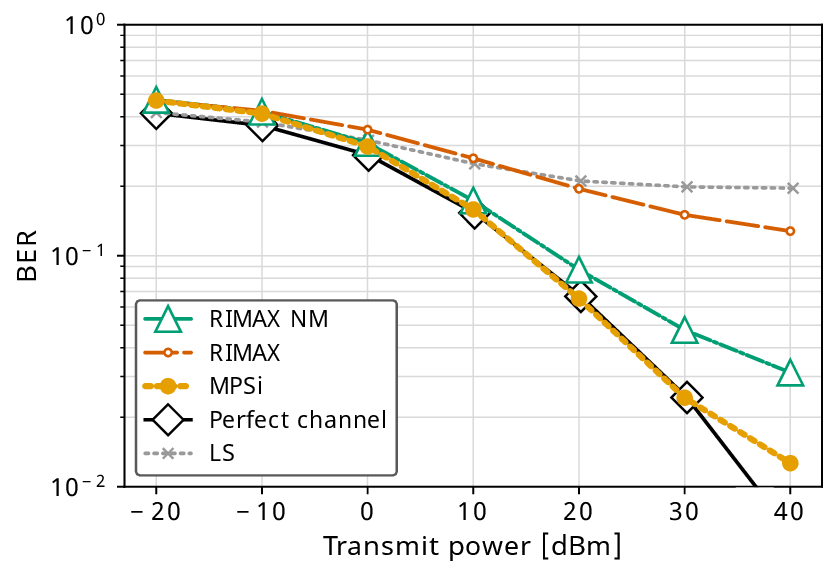}
			\caption{}
			\label{fig:Dundas_ber}
		\end{subfigure}
		\caption{BER versus pilot transmit power for single-user MRC detection: (a) Bahen, (b) Dundas.}
		\label{fig:BER_performance}
	\end{figure*}

	The resulting BER performance is shown in Fig.~\ref{fig:BER_performance}. MPSi approaches the perfect-channel benchmark in both the Bahen and Dundas scenarios, consistent with its low channel-estimation NMSE, whereas the RIMAX–NM variant falls short in the dense urban area of Dundas. RIMAX without NM initialization and the LS baseline exhibit higher BER in the reported configurations. Overall, MPSi achieves comparable BER while using a smaller, fixed refinement depth than the observed RIMAX iteration count.
	
	Note that MPSi uses pilot observations on 64 and 256 of the 1024 subcarriers in the Bahen and Dundas scenarios, respectively. These values correspond to pilot fractions of $6.25\%$ and $25\%$, both lower than the dense comb-type reference considered in~\cite{etsi.138.211:v18.2.0}. Thus, MPSi reduces pilot overhead and leaves a larger fraction of the available subcarriers for data transmission. Combined with the accuracy of its channel reconstruction, this reduced overhead can improve overall communication performance.
	
	\subsection{MPSi as a Model-Based Data Source for Wireless Learning}
	
	As a final application, we demonstrate how MPSi can generate pseudo-labels for the training of wireless learning systems. In practice, training datasets are often collected as received pilot measurements without corresponding ground-truth channel responses. Although such measurements can support self-supervised reconstruction objectives, they do not directly provide the target channels required for supervised tasks such as channel prediction or beam selection (e.g.,~\cite{HeathCodebook,physicsNn1}). Moreover, applying continual learning concepts to dynamic wireless environments~\cite{CLCioff,CLearning} requires a continuous stream of true channel labels that reflect the current propagation conditions. Obtaining such labels traditionally relies on ray tracing~\cite{physicsNn1}, which is computationally expensive and demands precise, up-to-date maps of all stationary and moving scatterers along with their material properties, making it a highly complex process.
	
	MPSi addresses this bottleneck by generating physically interpretable candidate labels directly from received pilots and a known sounding sequence. This workflow is particularly beneficial when site-specific channel state information is unavailable: a localized MPSi model can process measured pilots to produce high-fidelity environment estimates, which can then supervise downstream ML-enhanced algorithms for repeated training and inference within the deployment area.
	
\begin{figure*}[t]
		\begin{subfigure}[b]{0.49\textwidth}
			\centering
			\includegraphics[width=0.7\textwidth]{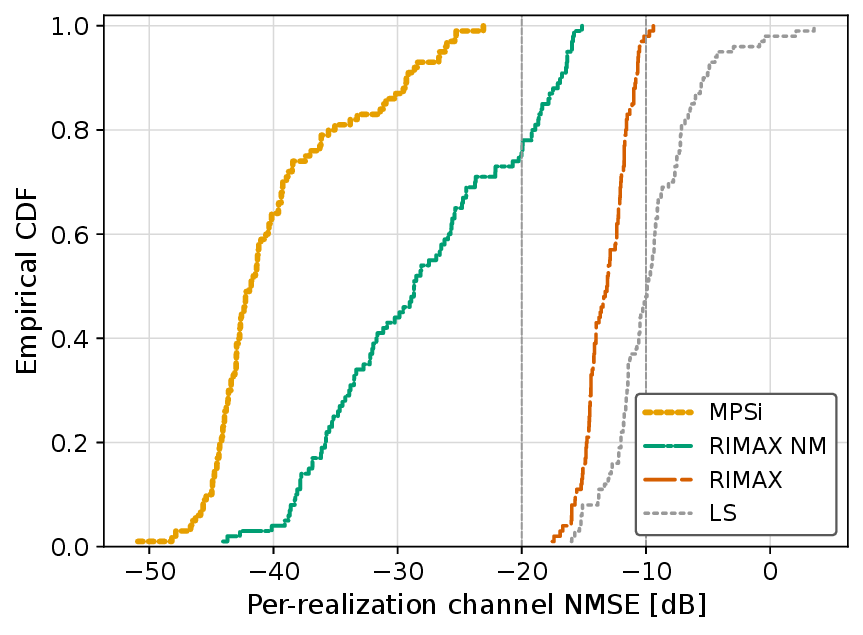}
			\caption{}
			\label{fig:Bahen_NMSE_CDF}
		\end{subfigure}
		\hfill
		\begin{subfigure}[b]{0.49\textwidth}
			\centering
			\includegraphics[width=0.7\textwidth]{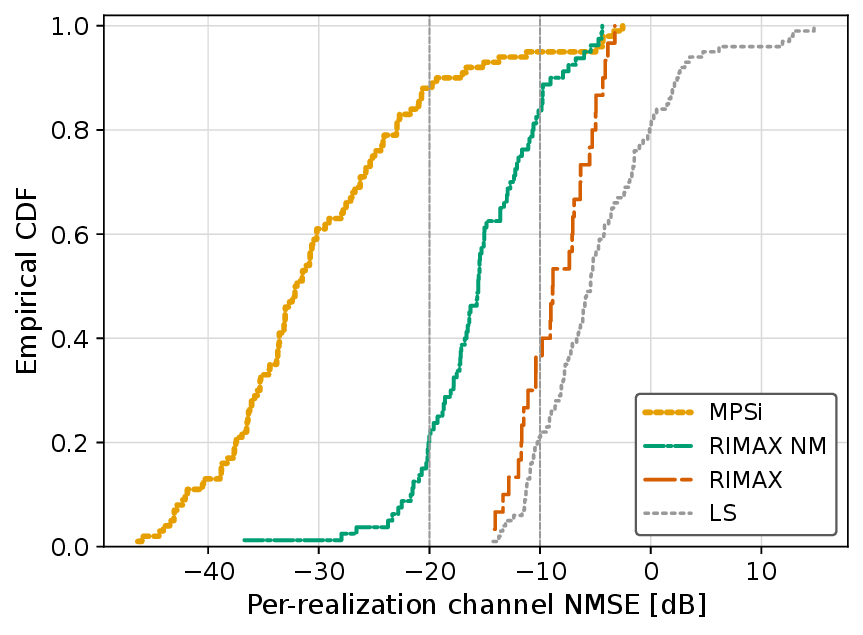}
			\caption{}
			\label{fig:Dundas_NMSE_CDF}
		\end{subfigure}
		\caption{Empirical CDF of the per-realization channel NMSE: (a) Bahen at $P_T=10$~dBm and (b) Dundas at $P_T=30$~dBm.}
		\label{fig:NMSE_CDF}
	\end{figure*}

	To evaluate the fidelity of the generated pseudo-labels, Fig.~\ref{fig:NMSE_CDF} shows the empirical CDF of the per-realization full-band channel NMSE in the Bahen and Dundas scenarios. The pilot transmit powers are set to $P_T=10$~dBm for Bahen and $P_T=30$~dBm for Dundas, respectively. In the Bahen scenario, all realizations have an NMSE below $-30$~dB, while more than $80\%$ of the realizations in the denser Dundas scenario have an NMSE below $-20$~dB. These results indicate that MPSi can produce accurate channel estimates across the evaluated site-specific environments and, under these configurations, achieves lower error than the RIMAX baselines.
	
\begin{figure*}[t]
		\begin{subfigure}[b]{0.49\textwidth}
			\centering
			\includegraphics[width=0.7\textwidth]{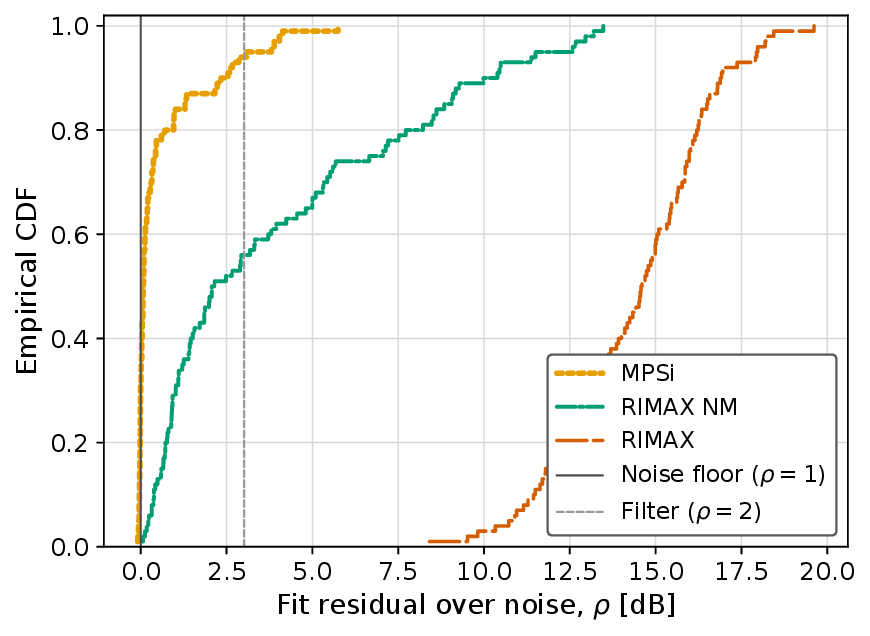}
			\caption{}
			\label{fig:Bahen_residual_CDF}
		\end{subfigure}
		\hfill
		\begin{subfigure}[b]{0.49\textwidth}
			\centering
			\includegraphics[width=0.7\textwidth]{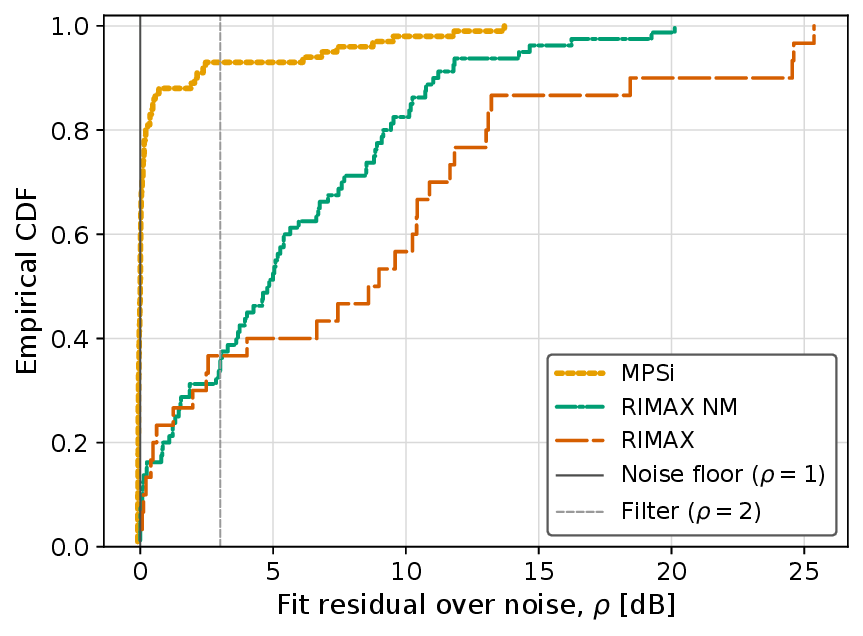}
			\caption{}
			\label{fig:Dundas_residual_CDF}
		\end{subfigure}
		\caption{Empirical CDF of the received-signal fit residual normalized by the realized noise energy for (a) Bahen and (b) Dundas.}
		\label{fig:residual_CDF}
	\end{figure*}

	The received-signal fit quality is evaluated in Fig.~\ref{fig:residual_CDF} using the normalized residual\ie 
	\begin{equation}
		\rho = \frac{\left\|\mathbf{Y}_{\mathrm{pilot}}-\widehat{\mathbf{Y}}_{\mathrm{pilot}}\right\|_F^2}{\left\|\mathbf{Y}_{\mathrm{pilot}}-\mathbf{Y}_{\mathrm{true}}\right\|_F^2},
	\end{equation}
	where $\widehat{\mathbf{Y}}_{\mathrm{pilot}}$ is the received signal reconstructed from the estimated paths and $\mathbf{Y}_{\mathrm{true}}$ is the corresponding noiseless received signal. Note that the residual analysis is an offline validation measure because the denominator uses the realized noise energy obtained from the known noiseless simulation signal. In a practical deployment, the same filter would require a calibrated or independently estimated noise level rather than access to $\mathbf{Y}_{\mathrm{true}}$. The horizontal axis in Fig.~\ref{fig:residual_CDF} is $\rho$, expressed in dB. Thus, $\rho=1$ (0~dB) corresponds to a reconstruction residual equal to the realized noise energy indicating that the estimated channel explains the observation down to the noise floor. Further, $\rho=2$ corresponds to the filtering threshold used in the analysis. We use $\rho_{\mathrm{th}}=2$ as a practical initial quality-control threshold, rather than as a universal statistical limit. This choice permits the residual to be at most twice the noise energy, approximately $3$~dB above the noise floor. This tolerance allows for moderate model mismatch while rejecting substantially poorer fits.
	
	As depicted in Fig.~\ref{fig:residual_CDF}, MPSi is concentrated close to the noise floor in both scenarios, whereas the RIMAX baselines exhibit larger residuals. This agreement between low channel NMSE and a residual close to the noise floor supports the use of MPSi estimates as candidate pseudo-labels. 

	\begin{table}[t]
\centering
\scriptsize
\setlength{\tabcolsep}{3pt}
\caption{Quality-controlled candidate labels.}
\label{tab:label_residual_filter}
\begin{tabular}{llrrr}
\toprule
Map & Method & \makecell{Retained\\fraction} & \makecell{Median\\NMSE kept} & \makecell{Worst\\NMSE kept} \\
\midrule
Bahen & MPSi & $1.00$ & $-39.2$ & $-31.8$ \\
 & RIMAX NM & $0.56$ & $-34.3$ & $-25.6$ \\
\midrule
Dundas & MPSi & $0.89$ & $-33.4$ & $-19.4$ \\
 & RIMAX NM & $0.29$ & $-18.6$ & $-9.7$ \\
\bottomrule
\end{tabular}
\end{table}

	Table~\ref{tab:label_residual_filter} summarizes the offline quality-control results. A candidate label is retained when its fit residual satisfies $\rho\leq2$ and its pilot-estimated signal-to-noise ratio (SNR) is at least $0$~dB. MPSi retains all Bahen realizations and $89\%$ of the Dundas realizations, with median retained-label NMSE values of $-39.2$ and $-33.4$~dB, respectively. RIMAX-NM retains fewer realizations and yields higher retained-label NMSE in both environments. These results support using MPSi estimates as candidate pseudo-labels, while the retained fraction and retained-label NMSE quantify label quality.

	
	\section{Conclusion}\label{sec:conclusion}
	
	In this paper, we introduced MPSi, a physics-guided neural framework to estimate the effective number and parameters of active propagation paths from received pilot observations and known transmitted pilot power. MPSi combines a differentiable path-count network, soft-top-$k$ gating, and fixed-depth unfolded Levenberg-Marquardt refinement while preserving a physically interpretable channel model. The resulting path-level representation supports initial access, wideband channel reconstruction, and the generation of candidate pseudo-labels for wireless-learning systems. Across the evaluated Bahen and Dundas ray-traced scenarios, MPSi achieved low channel NMSE and BER close to the perfect-CSI reference. Its initial-access performance was comparable to classical channel sounding algorithms\eg RIMAX, while its fixed refinement depth provided predictable control-flow complexity and substantially lower modeled air-interface overhead than dense codebook search. The residual-based quality-control analysis further showed how MPSi estimates can be screened as candidate pseudo-labels when measured pilots are available but perfect channel knowledge is not.
	
	These conclusions are limited to the two site-specific ray-traced environments representing small-cell and dense-urban scenarios. Possible future works include evaluating a broader range of environments and larger test sets, measuring platform-specific runtime, investigating MPSi-generated labels in downstream learning tasks, and extend the path model to include Doppler for mobile scenarios and channel prediction.
	

	\bibliographystyle{IEEEtran}
	\bibliography{references}

\end{document}